\pdfoutput=1
\documentclass[
    affil-it, % affiliations in italic
    auth-lg, % authors in large
    british, % dates in British format
    compression, % compressed title page
    contents, % show table of contents
    custom-packages={}, % Load listed custom packages
    references={bigone}, % Name of bibliography
]{lib/preprint}

\makecommand{sf}{\mathsf}{{Aut,ev,Sh,TD}}
\makecommand{fr}{\mathfrak}{{sugra}}

\begin{document}
    
    % Hypersetup
    \hypersetup{
        pdftitle = {Higher Gauge Theory},
        pdfauthor = {Leron Borsten,Mehran Jalali Farahani,Branislav Jurco,Hyungrok Kim,Jiri Narozny,Dominik Rist,Christian Saemann,Martin Wolf},
        pdfkeywords = {},
    }
    
    % Date of preprint
    \date{\today}
    
    % All emails in the order of appearance
    \email{l.borsten@herts.ac.uk,mj2020@hw.ac.uk,branislav.jurco@mff.cuni.cz,h.kim2@herts.ac.uk,narozny@karlin.mff.cuni.cz,d.rist@hw.ac.uk,c.saemann@hw.ac.uk,m.wolf @surrey.ac.uk}
    
    % All preprint numbers in the order of appearance
    \preprint{EMPG--24--01,DMUS--MP--24/01}
    
    % Title
    \title{Higher Gauge Theory} 
    
    % All authors
    \author[a]{Leron~Borsten\,\orcidlink{0000-0001-9008-7725}\,}
    \author[b]{Mehran~Jalali~Farahani\,\orcidlink{0009-0002-8282-9316}\,}
    \author[c]{Branislav~Jur{\v c}o\,\orcidlink{0000-0001-7782-2326}\,}
    \author[a]{Hyungrok~Kim\,\orcidlink{0000-0001-7909-4510}\,}
    \author[c]{Ji\v{r}\'{i}~N\'{a}ro\v{z}n\'{y}\,\orcidlink{0000-0003-3646-8070}\,}
    \author[b]{Dominik~Rist\,\orcidlink{0000-0002-1817-3458}\,}
    \author[b]{Christian~Saemann\,\orcidlink{0000-0002-5273-3359}\,}
    \author[d]{Martin~Wolf\,\orcidlink{0009-0002-8192-3124}\,}
    
    % All affiliations
    \affil[a]{Department of Physics, Astronomy, and Mathematics,\\ University of Hertfordshire, Hatfield AL10 9AB, United Kingdom}
    \affil[b]{Maxwell Institute for Mathematical Sciences, Department of Mathematics,\\ Heriot--Watt University, Edinburgh EH14 4AS, United Kingdom}
    \affil[c]{Mathematical Institute, Faculty of Mathematics and Physics,\\ Charles University, Prague 186 75, Czech Republic}
    \affil[d]{School of Mathematics and Physics,\\ University of Surrey, Guildford GU2 7XH, United Kingdom}
    
    % Abstract
    \abstract{This is an invited survey article on higher gauge theory for the Encyclopedia of Mathematical Physics, 2nd edition. In particular, we provide a lightning introduction to higher structures and to the construction of the kinematical data of higher gauge theories, i.e.~connections on higher principal bundles. We also summarise the most important applications and dynamical principles that have appeared in the literature, and we close with comments on related areas.}
    
    % Acknowledgements
    \acknowledgements{D.R.~and C.S.~were supported by the Leverhulme Research Project Grant RPG-2018-329. B.J.~was supported by the GA\v CR Grant EXPRO 19-28628X.}
    
    % Data and Licence Statement
    \datalicencemanagement{No additional research data beyond the data presented and cited in this work are needed to validate the research findings in this work. For the purpose of open access, the authors have applied a Creative Commons Attribution (CC BY) licence to any Author Accepted Manuscript version arising.}
    
    % Body
    \begin{body}
        
        \section{Introduction}\label{sec:intro}
        
        A gauge theory is a dynamical principle, such as an equation of motion or even a field theory Lagrangian, for kinematic data that involves a connection on a principal fibre bundle. Gauge theories are a crucial ingredient in our current modern description of fundamental physics. Both Maxwell's equations and general relativity are gauge theories, and the standard model of elementary particles is a quantum gauge theory.
        
        Recall that a principal fibre bundle $P$ over some smooth manifold $M$ is a fibration $\pi:P\rightarrow M$ with a smooth fibre preserving action of a structure Lie group $\sfG$ on $P$, which is principal, i.e.~free and transitive. The purpose of a connection on a principal fibre bundle is to define a notion of parallel transport, i.e.~a map from the paths in $M$ to $\sfG$, which describes the change along the path of objects on $M$ carrying a pointwise $\sfG$-action. In a neighbourhood $U$ of a point $p$ in $M$, a connection can be described by a $1$-form on $U$ taking values in the Lie algebra of $\sfG$, and the assignment is given by the path-ordered exponential
        \begin{equation}
            \gamma\ \mapsto\ P\exp\int_\gamma A
        \end{equation}
        for all paths $\gamma$ in $U$. This assignment defines the parallel transport functor from the  path groupoid\footnote{That is, a category with invertible morphisms. In this case, the objects are the points in $M$, and the morphisms are the paths between them.} of points and paths on $M$ to the structure group $\sfG$, regarded as a one-object groupoid. This functor is essentially in one-to-one correspondence with connections on principal fibre bundles~\cite{Barrett:1991aj,Caetano:1993zf,Mackaay:2000ac,Schreiber:0705.0452}.
        
        In many supergravity theories, and hence also in string/M-theory, as well as in six-dimensional conformal field theories, one naturally encounters connection-like objects which are described by forms of higher degree. This suggests a higher-dimensional form of parallel transport of extended objects along surfaces, volumes, etc. However, the most direct generalisation  is problematic, as evident compatibility requirements force the governing structure group to be Abelian by the Eckmann--Hilton argument~\cite{Hilton:1962:227-255}, cf.~\cite{Baez:0511710,Baez:2010ya}. 
        
        Regarding  parallel transport as a functor, however, offers a solution. As already mentioned, conventional parallel transport amounts to a functor between a path groupoid and a group, regarded as a one-object groupoid. Correspondingly, non-Abelian two-dimensional parallel transport should be regarded as a \uline{higher} functor from the path $2$-groupoid, with surfaces bounded by paths as $2$-morphisms, to a \uline{categorified} or $2$-group, regarded as a one-object $2$-groupoid, cf.~\cite{Mackaay:2000ac,Schreiber:1303.4663}. There are then evident generalisations to higher parallel transports of arbitrary dimension.
        
        The action of the categorified group in this picture requires a categorified principal bundle, whose total space is a  higher groupoid carrying a higher connection. Just as a conventional connection locally corresponds to a Lie algebra-valued $1$-form, a higher connection is described by higher forms taking values in certain components of higher Lie algebras. These higher principal bundles are also called gerbes, with a principal $n$-bundle, an $n$-fold categorified principal bundle, corresponding to an $(n-1)$-gerbe.
        
        The definition of most of these mathematical structures is straightforward, once one accepts the basic principles of categorification. One subtlety, however, is the appropriate definition of a connection. The evident definition resulting from categorification,  used in much of the literature, is too restrictive for applications in supergravity. In particular, the notion of curvature (which determines the notion of gauge transformations or bundle isomorphisms), needs to be corrected, or \uline{adjusted}. Once this is done, the Bianchi identities in higher gauge theory correctly reproduce, e.g., those of supergravity. A failure to incorporate the adjustment, however, leads to connections that merely appear non-Abelian, but are, in fact, locally gauge equivalent to Abelian connections. The latter observation may have led to a bias against higher non-Abelian principal bundles and higher non-Abelian parallel transport that unfortunately is still wide-spread, slowing the adoption of this powerful mathematical framework. We hope that this article can contribute to removing the underlying misconceptions.
        
        In fact, ignoring the mathematical difficulties of higher principal bundles, physicists constructed what they needed directly, employing standard physical and mathematical consistency requirements. This has led to a plethora of non-trivial higher gauge theories that implicitly use adjusted connections, avoiding the no-go objections. Most prominent are the supergravity theories arising in the low-energy limit of string/M-theory, which usually involve e.g.~the Kalb--Ramond $B$-field, an Abelian $2$-form gauge potential that couples to the world-volume of a string, or the Abelian M-theory $3$-form gauge potential that couples to the world-volume of the  M2-brane. These supergravity theories often involve further higher $(p+1)$-form fields, which in the fully fledged string theory couple to $p$-branes. 
        
        The second most prominent case is that of six-dimensional superconformal field theories. In particular, the maximally supersymmetric $(2,0)$ theory has all the earmarks of a higher gauge theory, but no known dynamical principle. Indeed, there are arguments that no such principle may be formulated at the classical level and that this theory is purely quantum.
        
        Thirdly, we have higher Chern--Simons theories and Alexandrov--Kontsevich--Schwarz--Zaboronsky (AKSZ) models, which are both of mathematical and physical interest.
        
        Finally, there is a large collection of other theories, from higher Stueckelberg models to teleparallel gravity that provide interesting examples of higher gauge theories to be studied.
        
        \paragraph{Outline.} 
        This survey article starts with a concise summary about higher structures, explaining categorification, higher spaces and higher groupoids, followed by a summary of the basics of higher Lie groups and higher Lie algebras in~\cref{sec:higher_structures}. \Cref{ssec:kinematical_data} then explains the description of the kinematical data of a higher gauge theory, first at the local, infinitesimal level in terms of $L_\infty$-algebra-valued forms, and then introducing connections on higher principal bundle, both with a focus on adjustments. In~\cref{sec:dynamics}, we list a number of higher gauge theories and describe their features. In order to provide a wider context, we close with comments on selected areas related to higher gauge theory in \cref{sec:related_areas}.
        
        Our emphasis in this review is to give a bird's eye perspective on higher gauge theory with enough background to understand the principles at work and to provide a relatively comprehensive list of references to the literature. It is clear that it is virtually impossible not to miss important work here and there, and we apologise for any such omissions.
        
        \section{Higher structures}\label{sec:higher_structures}
        
        The kinematical data of a higher gauge theory is described by a connection on a higher principal bundle. The latter is a categorified version of an ordinary principal bundle with categorified total space, categorified structure group, and, in principle, categorified base space. This section presents a short overview of these mathematical preliminaries, focusing on categorification, models of higher categories, and higher groupoids, groups and algebras. For a review with a focus on applications to M-theory, see~\cite{Jurco:2019woz}.
        
        \subsection{Higher categories and higher groupoids}
        
        Higher categories~\cite{Baez:9705009,cheng2004higher,Leinster:0107188,Leinster:2003aa,Simpson:2011aa} are mathematical gadgets which come with objects, $1$-morphisms between objects, $2$-morphism between $1$-morphisms, etc. These $n$-morphisms, $n\in\IN$, can usually have ends consisting of $(n-1)$-morphisms, and often there are two of these, leading to a globular structure such as 
        \begin{equation}
            \tikzset{Rightarrow/.style={double equal sign distance,>={Implies},->},triple/.style={-,preaction={draw,Rightarrow}},quad/.style={preaction={draw,Rightarrow,shorten >=0pt},shorten >=1pt,-,double,double distance=0.2pt}}
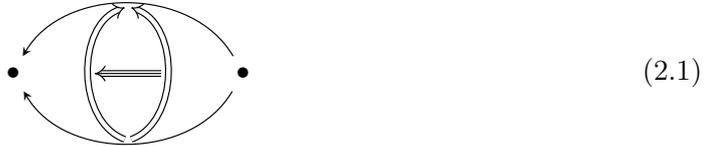
            \begin{tikzcd}[column sep=2.5cm,row sep=large]
                \bullet & \ar[l, bend left=60,  ""{name=U,inner sep=1pt,above}] \ar[l, bend right=60, , swap, ""{name=D,inner sep=1pt,below}] \bullet
                \arrow[Rightarrow, bend left=70,from=U, to=D,""{name=L,inner sep=1pt,right}]
                \arrow[Rightarrow, bend right=70,from=U, to=D,""{name=R,inner sep=1pt,left},swap]
                \arrow[triple,from=R,to=L]
            \end{tikzcd}
        \end{equation}
        Alternative arrangements exist, such as e.g.~identification of $n$-morphisms with simplicial $n$-simplices.
        
        \paragraph{Categorification.}
        The process of \uline{categorification}~\cite{Crane:1996yd,Baez:9802029} is not uniquely defined, contrary to the natural inverse process of \uline{decategorification} in which a category is reduced to the set of isomorphism classes of objects. This is mainly due to fact that there is a number of ways of turning a mathematical object into a higher one, and we briefly summarise the most important ones in the following.
        
        In many cases, the morphisms of a category do not only form a class but rather objects in an evident category, usually with a monoidal structure. For example, in any locally small category\footnote{i.e.~a category in which the morphisms between any two objects form a set}, we can replace the hom-sets with a corresponding monoidal category; the monoidal structure is needed to define composition as a morphism. This leads to the concept of an \uline{enriched category}, cf.~\cite{Kelly:2005}. For instance, a category enriched in $\CatCat$, the category of small categories\footnote{i.e.~locally small categories in which also the objects form a set} is a (strict) $2$-category, is an example of a higher category in which the $2$-morphisms are the natural transformations between functors in $\CatCat$. Iterating this process $n$ times then leads to an $n$-category. For instance, the iterative enrichment of the category of sets $\CatSet$ in $\CatCat$ yields $n\CatCat$, the category of strict (small) $n$-categories.
        
        Another process leading to a higher structure is \uline{horizontal categorification} or \uline{oidification}. Here, we regard a mathematical notion as a one-object category and extend this situation to a general category. Examples resulting from horizontal categorification are groupoids (oidified groups) and algebroids (oidified algebras).
        
        Then there is \uline{internalisation}. Here, we observe that any mathematical gadget is constructed from (i) a collection of sets, (ii) a collection of structure maps between these sets, and (iii) from a collection of structure relations or axioms satisfied by the structure maps. Given a category $\scC$, we can now replace these with (i) objects in a category, (ii) morphisms in a category, and (iii) relations between these morphisms. In this way, we can define e.g.~group objects in any category, and a group object in the category of topological spaces becomes a topological group, while a group object in the category of small categories becomes a strict $2$-category.
        
        There is also \uline{vertical categorification} which is less clearly defined. It can refer to an internalisation in $n\CatCat$, but one may choose to lift the strict relations between these $n$-functors by natural transformations and their higher variants. These additional lifting maps then need to satisfy coherence axioms in their own right. We hence have a choice of amount of strictness and level of coherence in our categorification, making it non-unique.
        
        Another useful procedure for categorification is to consider the internalisation of a simplicial set (i.e.~a collection of simplicial $n$-simplices, $n\in\IN$, with abstract simplicial face and degeneracy maps) internal to the category of a notion we want to categorify. For example, a simplicial object in the category of groups is a simplicial group, a model for an $\infty$-group. 
        
        \paragraph{Models for higher categories.} 
        There are many different models for higher categories. Most evidently, we have the objects in $n\CatCat$, the \uline{strict $n$-categories}. In such an $n$-category, we can compose $k$-morphisms in $n$ different directions\footnote{hence the name higher-dimensional algebra}. In the case of strict $2$-categories, e.g., we usually speak of horizontal and vertical composition. These compositions are associative and unital.
        
        A generalisation of this picture emerges after lifting unitality and associativity by filling the corresponding commutative diagrams of $k$-morphisms by particular $(k+1)$-morphisms, which then have to satisfy particular coherence axioms. We obtain weak $n$-categories, with the low-dimensional cases called \uline{bicategories}, \uline{tricategories}, and \uline{tetracategories}. Various semistrict versions exist, such as \uline{Gray categories}, which are semistrict tricategories.
        
        The coherence axioms in these become prohibitively complicated, and a good replacement is the notion of simplicial sets, in which each object that can be seen as a simplicial $n$-simplex with a missing face, a \uline{horn}, can be completed to a simplicial simplex. Depending on how generically this horn-filling condition is satisfied, one obtains \uline{weak Kan simplicial sets}, i.e.~\uline{quasi-categories}, or \uline{Kan simplicial sets}, i.e.~\uline{quasi-groupoids}. For more details, see e.g.~\cite{May:book:1967,joyal1999introduction,gabriel1967calculus,heuts2022simplicial}.
        
        There are many further models for higher categories, such as \uline{$n$-fold categories} (with \uline{double categories}, categories internal to $\CatCat$, for $n=2$), and there is a complicated web of relations between these that is still being explored.
        
        \paragraph{Higher groupoids.} 
        A \uline{higher groupoid} is a higher category in which all morphisms are equivalences. Explicit examples include \uline{bigroupoids} also known as \uline{weak $2$-groupoids}, which are weak $2$-categories in which the $1$-cells are weakly invertible (with respect to horizontal composition) and the $2$-cells are strictly invertible with respect to vertical composition. \uline{Quasi-groupoids}, i.e.~higher groupoids build on quasi-categories, are the same as \uline{Kan simplicial sets}, cf.~\cite{Zhu:0609420} or~\cite{Jurco:2016qwv} as a review.

        \paragraph{Lie groupoid equivalence.} 
        Lie groupoids have the problematic feature that fully faithful and essentially surjective functors between them fail to be equivalences. This is remedied by regarding Lie groupoids as objects in the weak $2$-category\footnote{This bicategory is equivalent to the bicategory of differentiable stacks~\cite{Pronk1996}.} $\CatBibun$ with bibundles (or Hilsum--Skandalis morphisms~\cite{MR925720}) as $1$-cells, cf.~\cite{Mrcun:1996aa,Metzler:0306176,Blohmann:2007ez,Lerman:0806.4160,Schommer-Pries:0911.2483}.
        
        In particular, a \uline{bibundle} $B:\scG\rightarrow\scH$ between two groupoids $\scG$ and $\scH$ is a smooth manifold $B$ with commuting left action of $\scG$ and right action of $\scH$ such that $B$ is a fibre bundle over the objects of $\scG$ together with a transitive and faithful action of $\scH$. A bibundle morphism between two bibundles $B,\tilde B:\scG\rightarrow\scH$ is an equivariant map $B\Rightarrow\tilde B$. Together, Lie groupoids, bibundles, and bibundle morphisms form the weak $2$-category $\CatBibun$~\cite{Pronk1996}. If $B$ can be regarded both as a bibundle $\scG\to\scH$ as well as $\scH\to\scG$, then we call $B$ a \uline{Morita} or \uline{weak equivalence} between $\scG$ and $\scH$. Note that any bibundle is equivalent to an anafunctor or a span of functors $\scG\overset{\cong}{\leftarrow}\hat\scG\rightarrow\scH$.
        
        An example of a relevant equivalence in $\CatBibun$ is the following.
        \begin{example}\label{exa:Cech_groupoid}        
            The \uline{\v Cech groupoid} of a surjective submersion $\sigma:Y\rightarrow M$ between smooth manifolds is the Lie groupoid 
            \begin{equation}
                \check\scC(\sigma)\ \coloneqq\ (Y^{[2]}\multirightarrow{2}Y)
                \ewith
                Y^{[2]}\ \coloneqq\ Y\times_MY~,
            \end{equation}
            with source and target maps $\sfs(y_1,y_2)\coloneqq y_2$ and $\sft(y_1,y_2)\coloneqq y_1$, identity map $\sfe(y)\coloneqq(y,y)$, composition $(y_1,y_2)\circ (y_2,y_3)\coloneqq(y_1,y_3)$, and inverse $(y_1,y_2)^{-1}=(y_2,y_1)$ for all $y\in Y$ and $(y_1,y_2),(y_2,y_3),(y_1,y_3)\in Y^{[2]}$.
            
            The \v Cech groupoid is Morita or weakly equivalent to $M$, trivially regarded as the discrete groupoid $(M\multirightarrow{2}M)$, with the total space of the bibundle given by $Y$ and evident projections and actions.
        \end{example}
        
        \paragraph{Higher spaces.} 
        Further generalising both spaces and higher groupoids are higher topoi, see~\cite{Lurie:0608040} or~\cite{Schreiber:2023nrr} in this volume for details. For an overview of higher geometry in physical applications, see also~\cite{Alfonsi:2023pps} in this volume.  
        
        Particularly interesting are \uline{cohesive $(\infty,1)$-topoi}~\cite{Schreiber:2013pra}, such as the \uline{$(\infty,1)$-topos $\CatGrpd_\infty$ of Kan simplicial sets}, the \uline{$(\infty,1)$-topos $\sfSh_{(\infty,1)}$ of higher smooth stacks}~\cite{carchedi2011categorical,ginot2013introduction,Blohmann:2007ez}, and its full sub-$(\infty,1)$-topos called the \uline{$(\infty,1)$-topos $\sfSh^\rmG_{(\infty,1)}$ of higher geometric stacks}~\cite{Toen:0404373,Pridham:0905.4044}. The latter is particularly interesting, as it allows for a formulation of higher Lie theory, see \cref{ssec:higher_Lie}. 
        
        \subsection{Higher groups}\label{sec:higher_groups}
        
        A \uline{higher group} is a higher groupoid with a single object, with a strict higher group being a strict higher groupoid with a single object, etc. 
        
        \begin{example}
            A \uline{weak $2$-group} is the same as the morphism category of a weak $2$-groupoid with a single object. This is a monoidal category $(\scG,\circ,\otimes)$, where $\circ$ and $\otimes$ denote vertical and horizontal composition, respectively, which is endowed with a unit object $\unit$, an inverse functor $-^{-1}:\scG\rightarrow\scG$, as well as four natural isomorphisms: the \uline{associator} $\sfa$, lifting associativity of horizontal composition, the \uline{left-} and \uline{right-unitors} $\sfl$ and $\sfr$, lifting the unity relations, and the \uline{unit} and \uline{counit} $\sfi$ and $\sfe$, lifting the inverse relations:
            \begin{equation}
                \begin{gathered}
                    \begin{tikzcd}[column sep=1.5cm]%,row sep=large]
                        \scG & 
                        \ar[l, bend left=45, "{(-\otimes-)\otimes-}", ""{name=U,inner sep=1pt,above}] \ar[l, bend right=45, "{-\otimes(-\otimes-)}", swap, ""{name=D,inner sep=1pt,below}] 
                        \arrow[Rightarrow,from=U, to=D, "{\sfa}",swap]
                        \scG\times \scG\times \scG
                    \end{tikzcd}
                    ~,~~~
                    \begin{tikzcd}[column sep=1.0cm]%,row sep=large]
                        \scG & 
                        \ar[l, bend left=45, "{-\otimes \unit}", ""{name=U,inner sep=1pt,above}] \ar[l, bend right=45, "{\sfid}", swap, ""{name=D,inner sep=1pt,below}] 
                        \arrow[Rightarrow,from=U, to=D, "{\sfl}",swap]
                        \scG
                    \end{tikzcd}
                    ~,~~~
                    \begin{tikzcd}[column sep=1.0cm]%,row sep=large]
                        \scG & 
                        \ar[l, bend left=45, "{\unit\otimes -}", ""{name=U,inner sep=1pt,above}] \ar[l, bend right=45, "{\sfid}", swap, ""{name=D,inner sep=1pt,below}] 
                        \arrow[Rightarrow,from=U, to=D, "{\sfr}",swap]
                        \scG
                    \end{tikzcd}
                    \\
                    \begin{tikzcd}[column sep=1.3cm]%,row sep=large]
                        \scG & 
                        \ar[l, bend left=45, "{\mathsf{pr}}", ""{name=U,inner sep=1pt,above}] \ar[l, bend right=45, "{(-\otimes -^{-1})\circ\Delta}", swap, ""{name=D,inner sep=1pt,below}] 
                        \arrow[Rightarrow,from=U, to=D, "{\sfi}",swap]
                        \scG
                    \end{tikzcd}
                    ~,~~~
                    \begin{tikzcd}[column sep=1.3cm]%,row sep=large]
                        \scG & 
                        \ar[l, bend left=45, "{(-^{-1}\otimes -)\circ\Delta}", ""{name=U,inner sep=1pt,above}] \ar[l, bend right=45, "{\mathsf{pr}}", swap, ""{name=D,inner sep=1pt,below}] 
                        \arrow[Rightarrow,from=U, to=D, "{\sfe}",swap]
                        \scG
                    \end{tikzcd}
                \end{gathered}
            \end{equation}
            In the above,
            $\mathsf{pr}$ is the projection functor onto $\unit$ and $\sfid_{\unit}$, and $\Delta:\scG\rightarrow\scG\times\scG$ is the diagonal map. The natural isomorphisms have to satisfy a number of coherence axioms, cf.~e.g.~\cite{Baez:0307200}.
            
            Weak $2$-groups form a weak $2$-category, and in this $2$-category, every weak $2$-group is equivalent to a strict $2$-group~\cite{Baez:0307200}, see also~\cite{ForresterBarker:2002aa} for a helpful review.
            
            Weak $2$-groups with trivial unitors, unit, and counit are sometimes called \uline{semistrict $2$-groups}~\cite{Jurco:2014mva}.
        \end{example}
        
        Another pathway to defining higher groups is to consider simplicial objects in the category of groups. The resulting \uline{simplicial groups} are convenient and useful models for higher groups. Simplicial groups contain large redundancies, and one can restrict to the defining information by considering the \uline{Moore complex}~\cite{Moore1954} of a simplicial group, which leads to the notion of a \uline{hypercrossed complex}~\cite{carrasco1986complejos,Conduche:1984:155,Carrasco:1991:195-235}. By the non-Abelian version~\cite{Conduche:1984:155}, see also~\cite{Carrasco:1991:195-235,Mutlu:9904038,Conduche:2003}, of the Dold--Kan correspondence~\cite{Dold:1961:201-312}, the set of simplicial $n$-simplices in a simplicial group is isomorphic to a semi-direct product of the elements of its Moore complex. Due to~\cite{Moore1954}, any simplicial group is automatically Kan. Furthermore, quasi-groupoids are the same as Kan simplicial sets, and quasi-groupoids with a single object are also called \uline{quasi-groups}. The category of quasi-groups is (Quillen) equivalent to the category of simplicial groups~\cite{quillen1969}.
        
        \begin{example}
            Simplicial groups that are constant from simplicial $1$-simplices onwards give rise to $1$-hypercrossed modules of groups, or, equivalently, crossed modules of groups~\cite{Conduche:1984:155}.
            
            A \uline{crossed module of groups} $(\sfH\xrightarrow{~\partial~}\sfG,\acton)$ is a pair of groups $\sfG$ and $\sfH$ together with a group morphism $\partial:\sfH\rightarrow \sfG$ as well as an action $\acton$ of $\sfG$ on $\sfH$ by automorphisms such that
            \begin{equation}
                \partial(g\acton h_1)\ =\ g\partial(h_1)g^{-1}
                \eand
                \partial(h_1)\acton h_2\ =\ h_1h_2h_1^{-1}
            \end{equation}
            for all $g\in\sfG$ and $h_{1,2}\in\sfH$.
            
            Crossed modules are objects in a weak $2$-category with butterflies and butterfly morphisms as $1$- and $2$-morphisms~\cite{Aldrovandi:0808.3627,Aldrovandi:0909.3350}.
        \end{example}
        
        \begin{example}
            Any strict $2$-group $(\scG_1\multirightarrow{2}\scG_0,\otimes,\circ)$ gives rise to a crossed module of groups $(\sfH\xrightarrow{~\partial~}\sfG,\acton)$ with 
            \begin{equation}
                \begin{gathered}
                    \sfG\ \coloneqq\ \scG_0~,~~~\sfH\ \coloneqq\ \ker(\sft)\ \subseteq\ \scG_1~,
                    \\
                    g_2g_1\ \coloneqq\ g_2\otimes g_1~,~~~h_2h_1\ \coloneqq\ h_2\circ(h_1\otimes\id_{\sfs(h_2)})~,
                    \\
                    \partial(h)\ \coloneqq\ \sfs(h^{-1})
                    \eand
                    g\acton h\ \coloneqq\ \id_g\otimes h\otimes\id_{g^{-1}}
                \end{gathered}
            \end{equation}
            for all $g_{1,2}\in \sfG$ and $h,h_{1,2}\in\sfH$. Conversely, any crossed module of groups $(\sfH\xrightarrow{~\partial~}\sfG,\acton)$ gives rise to a strict $2$-group $(\scG_1\multirightarrow{2}\scG_0,\otimes,\circ)$ with 
            \begin{equation}
                \begin{gathered}
                    \scG_0\ \coloneqq\ \sfG~,~~~\scG_1\ \coloneqq\ \sfH\rtimes\sfG~,
                    \\
                    (h_2,g_2)\otimes(h_1,g_1)\ \coloneqq\ ((g_2\acton h_1)h_2,g_2g_1)~,
                    \\
                    (h_2,g)\circ(h_1,\partial(h_2^{-1})g)\ \coloneqq\ (h_2h_1,g)
                \end{gathered}
            \end{equation}
            for all $g_{1,2}\in\sfG$ and $h_{1,2}\in\sfH$. Altogether, the weak $2$-categories of strict $2$-groups and crossed modules of groups are equivalent. We refer to~\cite{Brown:1976:296-302,Baez:0307200,Porst:0812.1464} for details.
        \end{example}
        
        \begin{example}\label{ex:2-group_BH}
            Let $\sfH$ be an Abelian group. Then, $(\sfH\xrightarrow{~\partial~}*,\acton)$ with $*$ the trivial group is a crossed module of groups.\footnote{The kernel of $\partial$ in a crossed module $(\sfH\xrightarrow{~\partial~}\sfG,\acton)$ is always an Abelian subgroup of $\sfH$ that is invariant under the $\sfG$-action $\acton$.} Further simple examples of crossed modules of groups are $(V\xrightarrow{~\partial~}\sfG,\acton)$, where $\partial$ is trivial, and $V$ is a representation of $\sfG$ with action $\acton$, as well as $(\sfG\xrightarrow{~\partial~}\sfAut(\sfG),\acton)$, where $\sfAut(\sfG)$ is the automorphism group of $\sfG$, the image of $\partial$ are the inner automorphisms and the action is evident. The latter $2$-group was used e.g.~in the first papers on non-Abelian gerbes with connections~\cite{Breen:math0106083,Aschieri:2003mw}.
        \end{example}
        
        Important examples of crossed modules of Lie groups, with the obvious definition, are the following.
        \begin{example}\label{ex:TDn}
            The crossed module of Lie groups $\sfTD_n$, which defines a categorical torus~\cite{Ganter:2014zoa} and describes geometric T-dualities\footnote{cf.~\cref{ssec:Tdualities}}~\cite{Nikolaus:2018qop}, is given by
            \begin{equation}\label{eq:def_TD_n}
                \begin{gathered}
                    \sfTD_n\ \coloneqq\ \big(\IZ^{2n}\times\sfU(1)\xrightarrow{~\partial~}\IR^{2n}\big)\,,
                    \\
                    \partial(m,\phi)\ \coloneqq\ m
                    ~,~~~
                    \xi\acton(m,\phi)\ \coloneqq\ (m,\phi-\inner{\xi}{m})
                \end{gathered}
            \end{equation}
            for all $\xi\in \IR^{2n}$, $m\in\IZ^{2n}$, and $\phi\in\sfU(1)$, where the groups are Abelian, and we used additive notation for $\sfU(1)$. The bilinear form is defined as 
            \begin{equation}
                \left\langle
                \begin{pmatrix}
                    v_1
                    \\
                    w_1
                \end{pmatrix},
                \begin{pmatrix}
                    v_2
                    \\
                    w_2
                \end{pmatrix}
                \right\rangle\ \coloneqq\ w_1 v_2
            \end{equation}
            for all $v_{1,2},w_{1,2}\in \IR^n$.
        \end{example}
        
        This is a finite-dimensional toy version of the following example.
        
        \begin{example}\label{ex:string_2_group}
            Let $\sfG$ be a compact simply-connected semi-simple Lie group and consider the crossed module of Lie groups
            \begin{equation}
                \sfString(\sfG)\ \coloneqq\ \big(\widehat{L_0\sfG}\xrightarrow{~\partial~}P_0\sfG,\acton\big)\,,
            \end{equation}
            where $P_0\sfG$ and $L_0\sfG$ are the based path and loop groups of $\sfG$ and $\widehat{L_0\sfG}$ is the Kac--Moody central extension of $L_0\sfG$, given by the normalised $2$-cocycle on this group. The products and actions are mostly the evident pointwise one, with the action of $P_0\sfG$ on the fibres of $\widehat{L_0\sfG}\rightarrow L_0\sfG$ more complicated and found in~\cite{Baez:2005sn}. When $\sfG=\sfSpin(n)$, this represents a $2$-group model for the \uline{string $2$-group $\sfString(n)$}~\cite{Baez:2005sn}\footnote{See~\cite{Rist:2022hci,Roberts:2022elp} for detailed formulas involving some corrections.}.
        \end{example}
        
        \subsection{Higher algebras and higher Lie theory}\label{ssec:higher_Lie}
        
        A weak categorification of the notion of a Lie algebra is a \uline{weak Lie $2$-algebra}, i.e.~a linear category, in which the bracket functor is antisymmetric and satisfies the Jacobi identity up to an \uline{alternator} and a \uline{Jacobiator}~\cite{Roytenberg:0712.3461}. In practice, one often restricts to hemistrict Lie $2$-algebras in which the Jacobiator is trivial, or, much more commonly, to semistrict Lie $2$-algebras, in which the alternator is trivial, see e.g.~the discussion in~\cite{Baez:2003aa,Borsten:2021ljb}.
        
        Just as in the case of $2$-groups, one can reduce a weak Lie $2$-algebra to a differential complex. In the weak, semistrict, and strict cases, one obtains a $2$-term \uline{$EL_\infty$-algebra}, a $2$-term \uline{$L_\infty$-algebra}, or a \uline{crossed module of Lie algebras}. The latter are simply infinitesimal versions of crossed modules of Lie groups, and they correspond to particular simplicial Lie algebras. More generally, simplicial Lie algebras give rise to hypercrossed complexes of Lie algebras, which describe \uline{strict higher Lie algebras}.
        
        Both $EL_\infty$-algebras\footnote{which have only been partially developed so far} and $L_\infty$-algebras are special instances of \uline{homotopy algebras}. For any algebra, one can construct a homotopy version by considering the Koszul resolution of the Koszul-dual of the operad of this algebra~\cite{Loday:2012aa}. In practice\footnote{with suitable assumptions on the finite-dimensionality of the underlying vector spaces, etc.}, this implies that the dual of a homotopy algebra is given by a semi-free\footnote{i.e.~free up to relations involving the differential} differential graded version of the Koszul-dual of the operad of the original algebra. 
        
        For example, the operad $\opLie$ of graded Lie algebras is Koszul-dual to the operad $\opCom$ of graded commutative associative and not necessarily unital algebras, and hence a homotopy Lie or $L_\infty$-algebra $\frL$ is the dual of a semi-free differential graded commutative algebra, the \uline{Chevalley--Eilenberg algebra} $\sfCE(\frL)$ of $\frL$. Given a $\IZ$-graded vector space $\frL$, we construct a free graded commutative algebra\footnote{Here, $\frL[i]$ for $i\in\IZ$ has homogeneously graded subspaces $(\frL[i])_k\coloneqq\frL_{k+i}$ for all $k\in\IZ$.} $\bigodot^\bullet(\frL[1])^*$. A differential on this algebra, i.e.~a map $Q_\sfCE:\bigodot^\bullet(\frL[1])^*\rightarrow\bigodot^\bullet(\frL[1])^*$ of degree~$1$ satisfying $Q_\sfCE^2=0$, and the evident graded Leibniz rule is fixed by its action on a basis of $(\frL[1])^*$, and these actions are dual to higher products $\mu_k:\frL^{\wedge k}\rightarrow\frL$ of degree $|\mu_k|=2-k$ for $k\in\IN$ satisfying the homotopy Jacobi identity
        \begin{equation}
            \sum_{j+k=i}\sum_{\sigma\in \text{Sh}(j;i) }\chi(\sigma;\ell_1,\ldots,\ell_{i})(-1)^{k}\mu_{k+1}(\mu_j(\ell_{\sigma(1)},\ldots,\ell_{\sigma(j)}),\ell_{\sigma(j+1)},\ldots,\ell_{\sigma(i)})\ =\ 0
        \end{equation}
        for all $\ell_{1,\ldots,i}\in\frL$. Here, $\text{Sh}(j;i)$ denotes $(j;i)$-unshuffles, i.e.~permutations with the first $j$ and the last $i-j$ elements ordered, and $\chi(\sigma;\ell_1,\ldots,\ell_i)$ is the Koszul sign defined by
        \begin{equation}
            \ell_1\wedge\ldots\wedge\ell_i\ =\ \chi(\sigma;\ell_1,\ldots,\ell_i)\,\ell_{\sigma(1)}\wedge\ldots\wedge\ell_{\sigma(i)}
        \end{equation}
        for all $\ell_{1,\ldots,i}\in\frL$. For a very detailed discussion, see e.g.~\cite{Jurco:2018sby,Jurco:2019bvp}.
        
        \begin{example}\label{exe:CEOrdinaryLieAlgebra}
            Consider a (finite-dimensional) Lie algebra $\frg$ with structure constants $\ttf^\alpha_{\beta\gamma}$ relative to a basis $\tte_\alpha$ with dual $\tte^\alpha$ on $(\frg[1])^*$, we have a differential and a dual Lie bracket
            \begin{equation}
                Q_\sfCE\tte^\alpha\ =\ -\tfrac12\ttf^\alpha_{\beta\gamma}\tte^\beta\tte^\gamma
                \eand
                [\tte_\alpha,\tte_\beta]\ =\ \ttf^\gamma_{\alpha\beta}\tte_\gamma
            \end{equation}
            for $\ttf^\alpha_{\beta\gamma}$ some constants, and $Q_\sfCE^2=0$ is equivalent to the Jacobi identity for the Lie bracket.
        \end{example}
        
        Importantly, $L_\infty$-algebras that are concentrated (i.e.~non-trivial exclusively) in non-positive degrees are \uline{higher Lie algebras} which are neither strict nor fully weak. The weakest form would be a yet to be developed, fully fledged notion of $EL_\infty$-algebras.
        
        \begin{example}\label{ex:string_Lie_2_algebra}
            Let $\frg$ be a real compact semi-simple metric Lie algebra. The archetypal example of a higher Lie algebra is the $2$-term $L_\infty$-algebra $\frstring(\frg)\coloneqq\IR\oplus\frg$ with the only non-trivial brackets
            \begin{equation}
                \begin{gathered}
                    \mu_2(-,-)\ \coloneqq\ [-,-]\,:\,\frg^{\wedge2}\ \rightarrow\ \frg~,
                    \\
                    \mu_3(-,-,-)\ \coloneqq\ \inner{-}{[-,-]}\,:\,\frg^{\wedge3}\ \rightarrow\ \IR~,
                \end{gathered}
            \end{equation}
            where $[-,-]$ is the Lie bracket and $\inner{-}{-}$ the metric on $\frg$, respectively, where $\inner{-}{-}$ is normalised such that $\inner{-}{[-,-]}$ generates $H^3(\frg,\IR)$. When $\frg=\frspin(n)$, this represents an $L_\infty$-algebra model of the \uline{string Lie algebra $\frstring(n)$}, cf.~\cite{Baez:2005sn}.
        \end{example}
        
        There is a \uline{higher version of Lie theory} which establishes a relation between higher Lie groups and higher Lie algebras, generalising that between Lie groups and Lie algebras. This generalisation is evident in the case of simplicial Lie groups and simplicial Lie algebras, which further translates to a Lie theory between hypercrossed complexes of Lie groups and Lie algebras. A general form of higher Lie theory can be formulated in the $(\infty,1)$-topos $\sfSh^\rmG_{(\infty,1)}$ of higher geometric stacks. In particular, a concrete and very useful perspective of Lie differentiation has been developed in~\cite{Severa:2006aa}, see also~\cite{Jurco20122389,Li:2014,Jurco:2014mva,Jurco:2016qwv,Li:2309.00901} for details. Here, quasi-groupoids are differentiated to $L_\infty$-algebroids, but the construction extends also e.g.~to Lie bigroupoids~\cite{Jurco:2014mva}. There are procedures for integrating $L_\infty$-algebras~\cite{Getzler:0404003,Henriques:2006aa,Severa:1506.04898}, but they often produce very large results in the expected weak isomorphism class~\cite{Sheng:1109.4002}.
        
        \begin{example}
            The application of the Lie functor to the crossed module of Lie groups from \cref{ex:string_2_group} yields the crossed module of Lie algebras
            \begin{equation}
                \big(L_0\frg\oplus\IR\xrightarrow{~\partial~}P_0\frg,\acton\big)\,.
            \end{equation}
            As an $L_\infty$-algebra, this crossed module of Lie algebras is quasi-isomorphic to the $L_\infty$-algebra from \cref{ex:string_Lie_2_algebra}, cf.~\cite{Baez:2005sn}.
        \end{example}
        
        \section{Kinematics: higher principal bundles with connections}\label{ssec:kinematical_data}
        
        \subsection{Locally and infinitesimally: the Becchi--Rouet--Stora--Tyutin complex}\label{ssec:BRST-complex}
        
        Much of the physics literature works with $\IR^{p,q}$ as a base space (often called spacetime) with connections given by differential forms on $\IR^{p,q}$ and gauge transformations discussed infinitesimally. This gives rise to the \uline{Becchi--Rouet--Stora--Tyutin (BRST) complex}, the Chevalley--Eilenberg resolution of the quotient of gauge potentials by gauge transformations, cf.~e.g.~\cite{Henneaux:1992,Mnev:2017oko,Jurco:2018sby} for reviews.
        
        Generally, one considers an action of a higher Lie algebra on a graded vector space of fields, and the BRST complex is the Chevalley--Eilenberg algebra of the corresponding action Lie algebroid. Concretely, consider a higher Lie algebra in the form of an $L_\infty$-algebra $\frL$ concentrated in non-positive degrees and a smooth manifold $M$. Let $U\subseteq M$ be open. A \uline{(local) $\frL$-valued connection} is a degree~$0$ morphism of graded commutative algebras 
        \begin{equation}\label{eq:degree0morphismCEOmega}
            \scA_0\,:\,\sfCE(\frL)\ \rightarrow\ \Omega^\bullet(U)~,
        \end{equation}
        where $\sfCE(\frL)$ is the Chevalley--Eilenberg algebra of $\frL$ and $\Omega^\bullet(U)$ the algebra of differential forms on $U$; cf.~\cite{Cartan:1949aaa,Cartan:1949aab,D'Auria:1982nx,Castellani:1991et,Bojowald:0406445,Kotov:2007nr,Sati:2008eg,Sati:2009ic,Fiorenza:2010mh,Gruetzmann:2014ica} for original references regarding this idea, as well as \cref{ssec:AKSZ} on the AKSZ model. The failure of $\scA_0$ to respect the differentials $Q_\sfCE$ on $\sfCE(\frL)$ and $\rmd$ (the exterior derivative) on $\Omega^\bullet(U)$, i.e.~the failure of $\scA_0$ to be a morphism of differential graded commutative algebras,  is the \uline{curvature}. In turn, its failure to be closed is described by the \uline{Bianchi identity}. Local and infinitesimal \uline{gauge transformations} are partially flat homotopies between two such morphisms. Together, the local expressions for the connection, the curvature, the Bianchi identity  and the gauge transformations form the \uline{kinematical data of a higher gauge theory}. The BRST complex is then the differential graded commutative algebra of morphisms of graded commutative algebras
        \begin{subequations}\label{eq:BRSTComplex}
            \begin{equation}\label{eq:morphismCEOmega}
                \scA_\bullet\,:\,\sfCE(\frL)\ \rightarrow\ \Omega^\bullet(U)
            \end{equation}
            of arbitrary but non-positive degrees (the inverse of the \uline{ghost degree}), equipped with the natural differential 
            \begin{equation}\label{eq:morphismCEOmegaDifferential}
                Q_{\rm BRST}\scA_\bullet\ \coloneqq\ \rmd\circ\scA_\bullet-\scA_\bullet\circ Q_\sfCE~.
            \end{equation}
        \end{subequations}
        
        \begin{example}\label{exa:morphismCEOmegaLieAlgebra}
            Consider a Lie algebra $\frg$ with the associated Chevalley--Eilenberg algebra as discussed in \cref{exe:CEOrdinaryLieAlgebra}. A $\frg$-valued connection $1$-form on $U\subseteq M$ is now defined by the image of the basis elements $\tte^\alpha$ of $(\frg[1])^*$ of a degree~$0$ morphism $\scA_0:\sfCE(\frg)\rightarrow\Omega^\bullet(U)$. In particular, we obtain this $1$-form by setting 
            \begin{equation}
                A\ \coloneqq\ \underbrace{\scA_0(\tte^\alpha)}_{\eqqcolon\,A^\alpha}\otimes\tte_\alpha\ \in\ \Omega^1(U)\otimes\frg~.
            \end{equation}
            Using the formulas in \cref{exe:CEOrdinaryLieAlgebra}, the curvature is then given by
            \begin{equation}
                F^\alpha\ \coloneqq\ (\rmd\circ\scA_0-\scA_0\circ Q_\sfCE)(\tte^\alpha)\ =\ \rmd A^\alpha+\tfrac12\ttf^\alpha_{\beta\gamma}A^\alpha\wedge A^\gamma~,
            \end{equation}
            i.e.~$F=\rmd A+\frac12[A,A]$ with $F\coloneqq F^\alpha\otimes\tte_\alpha\in\Omega^2(U)\otimes\frg$. In turn, this yields the Bianchi identity $\rmd F+[A,F]=0$. Infinitesimal gauge transformations are extracted from degree~$0$ morphisms $\tilde\scA_0:\sfCE(\frg)\rightarrow\Omega^\bullet(U\times[0,1])$ satisfying the partial flatness condition $\parder{t}\intprod\tilde F=0$ as
            \begin{equation}
                \delta A\ \coloneqq\ \left(\left.\parder{t}\tilde\scA_0(\tte^\alpha)\right|_{t=0}\right)\tte_\alpha\ =\ \rmd c+[A,c]~,
            \end{equation}
            where $c\coloneqq\parder{t}\intprod\tilde A\in\scC^\infty(U)\otimes\frg$ is the infinitesimal gauge parameter. 
            
            Generally, the morphisms~\eqref{eq:morphismCEOmega} defining the BRST complex have components of degree~$0$, the connection $1$-form $A^\alpha:(\frg[1])^*\rightarrow\Omega^1(U)$ itself, and of degree~$-1$, the \uline{ghost} $c^\alpha:(\frg[1])^*\rightarrow\scC^\infty(U)$. The differential~\eqref{eq:morphismCEOmegaDifferential} is then given by 
            \begin{equation}
                Q_{\rm BRST}c^\alpha\ =\ -\tfrac12\ttf^\alpha_{\beta\gamma}c^\beta c^\gamma
                \eand
                Q_{\rm BRST}A^\alpha\ =\ \rmd c^\alpha+\ttf^\alpha_{\beta\gamma}A^\beta c^\gamma~.
            \end{equation}
            
            The above constructions straightforwardly extend to general gauge $L_\infty$-algebras, with flat homotopies between homotopies encoding the higher gauge transformations, which requires the insertion of generators of higher ghost number in the BRST complex, cf.~e.g.~\cite{Jurco:2018sby}.
        \end{example}
        
        \subsection{Local fake flatness}\label{ssec:fake_flatness}
        
        One can lift the morphism $\scA_0$ in~\eqref{eq:degree0morphismCEOmega} of graded commutative algebras  to a morphism of differential graded commutative algebras $\tilde\scA_0$ by resolving its failure to respect the differential. To this end, we simply replace the Chevalley--Eilenberg algebra $\sfCE(\frL)$ with the \uline{Weil algebra} $\sfW(\frL)$, which is the Chevalley--Eilenberg algebra of the $L_\infty$-algebra of inner derivations, $T[1]\frL$, of $\frL$. Concretely, $\sfW(\frL)$ is the free graded commutative algebra generated by basis elements of $((T[1]\frL)[1])^*\cong (\frL[1])^*\oplus(\frL[2])^*$. We have the shift isomorphism $\sigma:(\frL[1])^*\rightarrow(\frL[2])^*$, and we extend $\sigma$ trivially by $\sigma((\frL[2])^*)\coloneqq 0$. The differential $Q_\sfW$ on $\sfW(\frL)$ is then defined by
        \begin{equation}\label{eq:actionQWeil}
            Q_\sfW\ \coloneqq\ Q_\sfCE+\sigma
            \eand
            Q_\sfW\circ\sigma+\sigma\circ Q_\sfCE\ =\ 0~,
        \end{equation}
        where $Q_\sfCE$ is the differential on $\sfCE(\frL)$. Now a \uline{(local) $\frL$-valued connection} on an open set $U\subseteq M$ of a smooth manifold $M$ is a degree~$0$ morphism of differential graded commutative algebras\footnote{This perspective becomes even more symmetric after noticing that $\Omega^\bullet(U)$ is the Weil algebra of $U\subseteq M$ regarded trivially as the Lie algebroid $(U\multirightarrow{2}U)$.}
        \begin{equation}
            \hat\scA_0\,:\,\sfW(\frL)\ \rightarrow\ \Omega^\bullet(U)~.
        \end{equation}
        
        \begin{example}
            Recall \cref{exa:morphismCEOmegaLieAlgebra}. In particular, for a Lie algebra $\frg$, the associated Weil algebra $\sfW(\frg)$ is generated by the basis vectors $\tte^\alpha$ of $(\frg[1])^*$ and the basis vectors $\hat\tte^\alpha\coloneqq\sigma(\tte^\alpha)$ of $(\frg[2])^*$, respectively. We define
            \begin{equation}
                A\ \coloneqq\ \underbrace{\hat\scA_0(\tte^\alpha)}_{\eqqcolon\,A^\alpha}\otimes\tte_\alpha\ \in\ \Omega^1(U)\otimes\frg
                \eand
                F\ \coloneqq\ \underbrace{\hat\scA_0(\hat\tte^\alpha)}_{\eqqcolon\,F^\alpha}\otimes\tte_\alpha\ \in\ \Omega^2(U)\otimes\frg~.
            \end{equation}
            Hence, using~\eqref{eq:actionQWeil} and the formulas from \cref{exa:morphismCEOmegaLieAlgebra}, we find
            \begin{equation}
                \begin{aligned}
                    \rmd A^\alpha\ &=\ (\rmd\circ\hat\scA_0)(\tte^\alpha)
                    \\
                    &=\ (\hat\scA_0\circ Q_\sfW)(\tte^\alpha)
                    \\
                    &=\ \hat\scA_0\big(-\tfrac12\ttf^\alpha_{\beta\gamma}\tte^\beta\tte^\gamma+\hat\tte^\alpha\big)
                    \\
                    &=\ -\tfrac12\ttf^\alpha_{\beta\gamma}A^\beta\wedge A^\gamma+F^\alpha~,
                \end{aligned}
            \end{equation}
            so that $F=\rmd A+\frac12[A,A]$. Likewise, we find the Bianchi identity $\rmd F+[A,F]=0$.
        \end{example}
        
        \begin{example}
            An early example of the above construction was given in the context of supergravity~\cite{DAuria:1980cmy,D'Auria:1982nx}, see also~\cite{Castellani:1991et} and~\cite{Fre:2008qw}, where the above picture is also known under the name of \uline{free differential algebra} (FDA). The modern description~\cite{Sati:2008eg} starts from the Lie $3$-algebra $\frsugra(1,10)$, which consists of a Lie $3$-algebra extension of the super Poincar\'e algebra in eleven dimensions by a super Lie algebra $4$-cocycle, analogously to the extension of $\frg$ by a Lie algebra $3$-cocycle to $\frstring(\frg)$, cf.~\cref{ex:string_Lie_2_algebra}. The Chevalley--Eilenberg generators of the super Poincar\'e algebra map to the vielbein, the spin connection, and the gravitino, and the additional generator in degree~$4$ maps to the supergravity $C$-field. The Weil extension then defines the corresponding curvature expressions: the torsion, the Riemann curvature, the covariant derivative of the gravitino as well as the adjusted $4$-form curvature. For the detailed discussion, see~\cite{Sati:2008eg}.
        \end{example}
        
        The \uline{Weil extension} of the BRST complex~\eqref{eq:BRSTComplex} is found by considering morphisms of graded commutative algebras
        \begin{subequations}\label{eq:WeilExtendedBRSTComplex}
            \begin{equation}
                \scA_\bullet\,:\,\sfW(\frL)\ \rightarrow\ \Omega^\bullet(U)
            \end{equation}
            of arbitrary but non-positive degrees and with the additional requirement that $\hat\scA_\bullet$ vanishes on the image of $\sigma$ for negative degrees, cf.~\cite{Saemann:2019dsl}. The latter ensures that gauge symmetries are not doubled, but it also leads to constraints on all but the top-form-degree component of the curvature, which are also called the \uline{fake-flatness conditions}. The differential~\eqref{eq:morphismCEOmegaDifferential} generalises as
            \begin{equation}\label{eq:WeilExtendedBRSTDifferential}
                Q_{\rm BRST}\scA_\bullet\ \coloneqq\ \rmd\circ\scA_\bullet-\scA_\bullet\circ Q_\sfW~.
            \end{equation}
        \end{subequations}
        
        \begin{example}\label{eq:WeilAlgebra2Term}
            As an illustrative example, consider a $2$-term $L_\infty$-algebra $\frL=\frL_{-1}\oplus\frL_0$ with generic Weil algebra generated by $\tte^\alpha=(\ttr^a,\ttt^i)$ and $\hat\tte^\alpha=(\hat\ttr^a,\hat \ttt^i)$ and differential defined by its action 
            \begin{equation}\label{eq:Weil_algebra_generic_Lie_2}
                \begin{aligned}
                    (Q_\sfW\ttr^a,Q_\sfW\ttt^i)\ &\coloneqq\ \big(\tfrac1{3!}\ttf^a_{ijk}\ttt^i\ttt^j\ttt^k-\ttf^a_{ib}\ttt^i\ttr^b+\hat\ttr^a,-\tfrac12\ttf^i_{jk}\ttt^j\ttt^k-\ttf^i_a\ttr^a+\hat\ttt^i\big)~,
                    \\
                    (Q_\sfW\hat\ttr^a,Q_\sfW\hat\ttt^i)\ &\coloneqq\ \big(-\tfrac12\ttf^a_{ijk}\ttt^i\ttt^j\hat\ttt^k+\ttf^a_{ib}\hat\ttt^i\ttr^b-\ttf^a_{ib}\ttt^i\hat\ttr^b,-\ttf^i_{jk}\ttt^j\hat\ttt^k+\ttf^i _a\hat\ttr^a\big)\,,
                \end{aligned}
            \end{equation}
            where $\ttf^i_{jk}$, $\ttf^i_a$, $\ttf^a_{ib}$, and $\ttf^a_{ijk}$ are some constants. The complex~\eqref{eq:WeilExtendedBRSTComplex} is then given by
            \begin{equation}
                \begin{gathered}
                    d\ \coloneqq\ \underbrace{\hat\scA_{-2}(\ttr^a)}_{\eqqcolon\,d^a}\otimes\ttr_a\ \in\ \scC^\infty(U)\otimes\frL_{-1}~,
                    \\
                    c\ \coloneqq\ \underbrace{\hat\scA_{-1}(\ttt^i)}_{\eqqcolon\,c^i}\otimes\ttt_i+\underbrace{\hat\scA_{-1}(\ttr^a)}_{\eqqcolon\,c^a}\otimes\ttr_a\ \in\ \scC^\infty(U)\otimes\frL_0\,\oplus\,\Omega^1(U)\otimes\frL_{-1}~,
                    \\
                    a\ \coloneqq\ \underbrace{\hat\scA_0(\ttt^i)}_{\eqqcolon\,A^i}\otimes\ttt_i+\underbrace{\hat\scA_0(\ttr^a)}_{\eqqcolon\,B^a}\otimes\ttr_a\ \in\ \Omega^1(U)\otimes\frL_0\,\oplus\,\Omega^2(U)\otimes\frL_{-1}~,
                    \\
                    f\ \coloneqq\ \underbrace{\hat\scA_0(\hat\ttt^i)}_{\eqqcolon\,F^i}\otimes\ttt_i+\underbrace{\hat\scA_0(\hat\ttr^a)}_{\eqqcolon\,H^a}\otimes\ttr_a\ \in\ \Omega^2(U)\otimes\frL_0\,\oplus\,\Omega^3(U)\otimes\frL_{-1}~.
                \end{gathered}
            \end{equation}
            Here, $d$ is the \uline{ghost-for-ghost}, $c$ the ghost, $a$ the connection, and $f$ the curvature. The fact that $\hat\scA_1$ is trivial implies the curvature expressions 
            \begin{equation}
                \begin{aligned}
                    F^i\ &=\ \rmd A^i+\tfrac12\ttf^i_{jk}A^j\wedge A^k+\ttf^i_aB^a~,
                    \\
                    H^a\ &=\ \rmd B^a+\ttf^a_{ib}A^i\wedge B^b-\tfrac1{3!}\ttf^a_{ijk}A^i\wedge A^j\wedge A^k
                \end{aligned}
            \end{equation}
            and the Bianchi identities
            \begin{equation}
                \begin{aligned}
                    \rmd F^i\ &=\ -\ttf^i_{jk}A^j\wedge F^k+\ttf^i_a H^a~,
                    \\
                    \rmd H^a\ &=\ -\ttf^a_{ib}A^i\wedge H^b+\ttf^a_{ib}F^i\wedge B^b-\tfrac12\ttf^a_{ijk}A^i\wedge A^j\wedge F^k~.
                \end{aligned}
            \end{equation}
            Gauge transformations are given by the actions of the BRST differential~\eqref{eq:WeilExtendedBRSTDifferential}. Furthermore, the triviality of $\hat\scA_{-1}(\hat\ttr^a)$ implies that
            \begin{equation}\label{eq:ex_constraint}
                -\tfrac12\ttf^a_{ijk}c^ic^jF^k+\ttf^a_{ib}F^id^b\ =\ 0
                \quad\Leftrightarrow\quad
                \ttf^a_{ijk}F^k\ =\ 0\ =\ \ttf^a_{ib}F^i~,
            \end{equation}
            and this shows that gauge transformations only close if $F^i$ satisfies a particular condition\footnote{Physicists say that the BRST complex is \uline{open}.}. Fake flatness here amounts to $F^i=0$, which is sufficient for satisfying~\eqref{eq:ex_constraint}. This condition is problematic, as it implies that locally, $A^i$ can always be gauged away~\cite{Gastel:2018joi,Saemann:2019dsl}, which renders this type of connection unsuitable in many physical applications. Analogous conditions to~\eqref{eq:ex_constraint} have been identified in the consistent definition of parallel transport~\cite{Baez:2004in,Schreiber:2008aa} as well as the gluing of finite gauge transformations for higher principal $2$-bundles with connection~\cite{Rist:2022hci}. In the former case, $F^i=0$ was identified as a necessary condition.
        \end{example}
        
        \subsection{Local adjustments}\label{ssec:local_adjustments}
        
        The undesirable consequences of fake flatness can be avoided by performing a coordinate change or deformation of the Weil algebra, which changes the notion of curvature and implies a change of gauge transformations and Bianchi identities~\cite{Sati:2008eg,Sati:2009ic,Saemann:2019dsl}. In particular, an \uline{adjustment} of an $L_\infty$-algebra $\frL$ is a deformation of its Weil algebra $\sfW(\frL)$ preserving the projection $\sfW(\frL)\rightarrow\sfCE(\frL)$ and leading to a closed Weil-extended BRST complex~\eqref{eq:WeilExtendedBRSTComplex}. The object giving rise to the deformation was called a \uline{Chern--Simons term} in~\cite{Sati:2008eg,Sati:2009ic}. 
        
        \begin{example}
            The archetypal example here is the definition of a connection for the $2$-term $L_\infty$-algebra $\frstring(\frg)=\IR\oplus\frg$ introduced in \cref{ex:string_Lie_2_algebra}. Following our discussion from \cref{eq:WeilAlgebra2Term}, the associated Weil algebra $\sfW(\frstring(\frg))$ has generators $\tte^\alpha=(\ttr,\ttt^i)$ and $\hat\tte^\alpha=(\hat\ttr,\hat\ttt^i)$ of degrees~$(2,1)$ and~$(3,2)$, respectively, and the differential
            \begin{equation}
                \begin{aligned}
                    (Q_\sfW\ttr,Q_\sfW\ttt^i)\ &\coloneqq\ \big(\tfrac1{3!}\ttf_{ijk}\ttt^i\ttt^j\ttt^k+\hat\ttr,-\tfrac12\ttf^i_{jk}\ttt^j\ttt^k+\hat\ttt^i\big)\,,
                    \\
                    (Q_\sfW\hat\ttr,Q_\sfW\hat\ttt^i)\ &\coloneqq\ \big(-\tfrac12\ttf_{ijk}\ttt^i\ttt^j\hat\ttt^k,-\ttf^i_{jk}\ttt^j\hat\ttt^k\big)\,,
                \end{aligned}
            \end{equation}
            where, as before, $\ttf^i_{jk}$ are the Lie algebra structure constants of $\frg$, $\kappa_{ij}$ are the structure constants of the metric $\inner{-}{-}$ on $\frg$, and we have set $\ttf_{ijk}\coloneqq\kappa_{il}\ttf^l_{jk}$. We now deform $\sfW(\frstring(\frg))$ to obtain a deformed Weil algebra $\tilde\sfW(\frstring(\frg))$ that has the same generators as $\sfW(\frstring(\frg))$ but with the differential given by
            \begin{equation}
                \begin{aligned}
                    (Q_{\tilde\sfW}\ttr,Q_{\tilde\sfW}\ttt^i)\ &\coloneqq\ \big(\tfrac1{3!}\ttf_{ijk}\ttt^i\ttt^j\ttt^k-\kappa_{ij}\ttt^i\hat\ttt^j+\hat\ttr,-\tfrac12\ttf^i_{jk}\ttt^j\ttt^k+\hat\ttt^i\big)\,,
                    \\
                    (Q_{\tilde\sfW}\hat\ttr,Q_{\tilde\sfW}\hat\ttt^i)\ &\coloneqq\ \big(\kappa_{ij}\hat\ttt^i\hat\ttt^j,-\ttf^i_{jk}\ttt^j\hat\ttt^k\big)\,.
                \end{aligned}
            \end{equation}
            Using $Q_{\tilde\sfW}$, the Weil-extended BRST complex~\eqref{eq:WeilExtendedBRSTComplex} then leads to a connection
            \begin{equation}
                a\ =\ A+B\ \in\ \Omega^1(U)\otimes\frg\,\oplus\,\Omega^2(U)
            \end{equation}
            with curvatures
            \begin{subequations}
                \begin{equation}
                    f\ =\ F+H\ \in\ \Omega^2(U)\otimes\frg\,\oplus\,\Omega^3(U)~,
                \end{equation}
                with 
                \begin{equation}
                    F\ \coloneqq\ \rmd A+\tfrac12[A,A]
                    \eand
                    H\ \coloneqq\ \rmd B+\inner{A}{\rmd A}+\tfrac13\inner{A}{[A,A]}
                \end{equation}
            \end{subequations}
            satisfying the Bianchi identities
            \begin{equation}\label{eq:Bianchi_identity_2_algebra}
                \rmd F+[A,F]\ =\ 0
                \eand 
                \rmd H\ =\ \inner{F}{F}~.
            \end{equation}
            Infinitesimal gauge transformations are parametrised by 
            \begin{equation}
                c+\Lambda\ \in\ \scC^\infty(U)\otimes\frg\,\oplus\,\Omega^1(U)
            \end{equation}
            and are given by 
            \begin{equation}
                \delta A\ =\ \rmd c+[A,c]
                \eand
                \delta B\ =\ \rmd\Lambda-\inner{c}{\rmd A}
            \end{equation}
            so that
            \begin{equation}
                \delta F\ =\ -[c,F]
                \eand
                \delta H\ =\ 0~.
            \end{equation}
            
            Such kinematical data naturally arises in the context of supergravity, see~\cref{ssec:tensor_hierarchies}, where also many more examples of adjustments arise naturally. In particular, there is a natural way of constructing an $L_\infty$-algebra from a shift-truncated differential graded Lie algebra~\cite{Fiorenza:0601312,Getzler:1010.5859}. A refinement of this process~\cite{Borsten:2021ljb} naturally leads to adjustment data.
        \end{example}
        
        \subsection{Higher principal bundles}\label{ssec:higher_principle_bundles}
        
        The notion of an action of a higher group $\scG$ on a higher groupoid $\scH$ is defined as an appropriate higher functor $\acton:\scG\rightarrow\sfAut(\scH)$, where $\sfAut(\scH)$ denotes the higher automorphism group of $\scH$.
        
        \paragraph{Higher principal bundles.}
        Consider a Lie $n$-group $\scG$. A \uline{principal $\scG$-bundle} over a smooth manifold $M$ is a Lie $(n-1)$-groupoid $\scP$ together with a surjective submersion functor $\pi:\scP\rightarrow M$ and a smooth right action of $\scG$ on $\scP$ that preserves $\pi$ such that the smooth functor
        \begin{equation}\label{eq:shear_map}
            \scP\times\scG\ \rightarrow\ \scP\times_M\scP
        \end{equation}
        is a weak groupoid equivalence. The automorphisms of $\scP$, providing bundle isomorphisms or gauge transformations, are then defined in the evident manner.
        
        Equivalently, there is a description of a principal $\scG$-bundle in terms of \uline{$\scG$-valued \v{C}ech cocycles}. Consider a smooth manifold $M$ together with a surjective submersion $\sigma:Y\rightarrow M$. As explained in \cref{exa:Cech_groupoid}, there is a weak equivalence between $M$, trivially regarded as a groupoid $(M\multirightarrow{2}M)$, and the \v{C}ech groupoid $\check\scC(\sigma)=(Y^{[2]}\multirightarrow{2}Y)$; see \cref{exa:Cech_groupoid}. The \v{C}ech cocycles are then sufficiently weak functors
        \begin{equation}
            \Gamma\,:\,\check\scC(\sigma)\ \rightarrow\ \sfB\scG~,
        \end{equation}
        where $\check\scC(\sigma)$ is trivially regarded as a higher groupoid and $\sfB\scG$ is the delooping of $\scG$, i.e.~the one-object higher groupoid that encodes the same operations as the higher group $\scG$. Gauge transformations or \uline{\v Cech coboundaries} are then given by higher natural isomorphisms between the corresponding functors. Higher gauge transformations are obtained from higher invertible transfors, starting with modifications of natural isomorphisms. 
        
        Finally, the total space of a principal $\scG$-bundle can be regarded as the homotopy pullback
        \begin{equation}
            \begin{tikzcd}
                & \scP\arrow[d]\arrow[d]\arrow[r] 
                & *\arrow[d]\arrow[d]
                \\
                & M \arrow[Rightarrow,ur]\arrow[r,"\hat\Gamma"'] 
                & B\scG
            \end{tikzcd}
        \end{equation}
        where $*\cong E\scG$ replaces the contractible universal bundle $E\scG$ over the classifying space $B\scG$. The map $\hat \Gamma$ can be identified with the precomposition of $\Gamma$ by the bibundle equivalence $M\rightarrow\check\scC(\sigma)$. 
        
        For further details on general principal $\infty$-bundles, see~\cite{Nikolaus:1207ab,Schreiber:1207.0249} as well as~\cite{Bunk:2023jsj} in this volume (and also~\cite{Jurco:2019bvp}) as well as the extensive~\cite{Sati:2021eyj}. A very helpful review of the different pictures of principal $2$-bundles is~\cite{Nikolaus:2011ag}.
        
        \paragraph{Presentation.}
        Even though the theory exposed above can be seen as being of $1$-categorial nature (higher morphisms are considered as homotopy maps, higher relations translate to $1$-category diagrams), the data with which it deals are sufficient for a reconstruction (\emph{presentation}) higher category objects. More concretely, we have chosen to work in the $(\infty,1)$-topos $\sfSh_{(\infty,1)}$ of higher smooth stacks; for details on the following, see~\cite{Nikolaus:1207ab,Schreiber:1207.0249} as well as~\cite{Bunk:2020wla} for a recent application. 
        
        More generally, such presentation is realised by $\CatsSh(\CatCart)$ the category of simplicial sheaves on Cartesian spaces. Let $\scM$ be an object in $\CatsSh(\CatCart)$ and $\scG$ a group object internal to $\CatsSh(\CatCart)$, respectively. The surjective submersion functor is then a local fibration in $\CatsSh(\CatCart)_{\rm lp}$ which is $\CatsSh(\CatCart)$ equipped with weak equivalences being stalkwise weak equivalences of simplicial sets and fibrations as stalkwise Kan fibrations of simplicial sets~\cite{Schreiber:1207.0249}. 

        There is now an equivalence of Kan simplicial sets,
        \begin{equation}
            \scG\sfBun(\scM)\ \cong\ \sfhom_{\CatsSh(\CatCart)}(\scM,\overline{\sfW}\scG)~,
        \end{equation}        
        where $\scG\sfBun(\scM)$ is the $\infty$-groupoid of principal $\scG$-bundles over $\scM$, and $\sfhom_{\CatsSh(\CatCart)}(\scM,\overline{\sfW}\scG)$ is the $\infty$-groupoid of cocycles from $\scM$ into the simplicial classifying space $\overline{\sfW}\scG$ of $\scG$. Here, $\overline{\sfW}$ is the classifying space functor, cf.~\cite{Porter:2007aa}. Essentially, $\overline{\sfW}\scG$ replaces the delooping $\sfB\scG$ of a higher group $\scG$ in the simplicial context, and its d\'ecalage, $\sfW\scG\coloneqq\sfDec(\overline{\sfW}\scG)$, which is a principal $\scG$-bundle over $\overline{\sfW}\scG$, replaces the universal bundle $E\scG\rightarrow B\scG$. The cocycles of a principal $\scG$-bundle $\scP$ over $\scM$ are then simplicial maps $g$ in
        \begin{equation}
            \begin{tikzcd}
                \scM & \ar[l,"\cong"']\scY\ar[r, "g"] & \overline{\sfW}\scG~,
            \end{tikzcd}
        \end{equation}
        where $\scY$ is a cofibrant replacement of a Kan simplicial manifold weakly equivalent to $\scM$. The total space of $\scP$ is obtained by pulling back $\sfW\scG$ along $g$ with the evident projection to $\scM$. For further details, see also~\cite{Bakovic:2009aa}.
        
        \paragraph{Nomenclature.}
        Higher principal bundles with a Lie $n$-group as structure group are also called \uline{principal $n$-bundles}. A \uline{gerbe}~\cite{Giraud:1971} is a principal $2$-bundle, usually with the structure $2$-group of \cref{ex:2-group_BH} and hence an \uline{Abelian gerbe}, see also~\cite{Moerdijk:0212266,0817647309,Husemller:2008aa}. A \uline{Hitchin--Chatterjee gerbe}~\cite{Hitchin:1999fh,Chatterjee:1998} is an Abelian gerbe defined in terms of cocycles subordinate to a cover. A \uline{bundle gerbe}~\cite{Murray:9407015,Murray:2007ps} is a more general notion of an Abelian gerbe that is more suitable for geometric constructions. The paper~\cite{Aschieri:2003mw} describes a non-Abelian generalisation of these bundle gerbes. We will use the term principal $n$-bundle to capture all of these structures.
        
        \begin{example}\label{exa:2Bundle}
            An instructive example is that of a principal $\scG$-bundle $\scP$ for $\scG$ a strict Lie $2$-group. The total space of $\scP$ is a Lie groupoid, together with a functor $\pi:\scP\rightarrow(M\multirightarrow{2}M)$ and an action of $\scG$ on $\scP$ preserving $\pi$ such that~\eqref{eq:shear_map} is an equivalence. This implies that there is a surjective submersion $\sigma:Y\rightarrow M$ such that there is an equivalence of Lie groupoids between $\scP$ and the tensor category $\check\scC(\sigma)\times\scG$, where $\check\scC(\sigma)$ is the \v Cech groupoid, cf.~\cref{exa:Cech_groupoid}. Details on this total space perspective are found in~\cite{Wockel:2008aa,Nikolaus:2011ag,Waldorf:1608.00401}. 
            
            Subordinate to a sufficiently fine surjection $\sigma:Y\rightarrow M$, the principal $\scG$-bundle $\scP$ is given by a weak $2$-functor $\hat\Gamma:\check\scC(\sigma)\rightarrow\sfB\scG$. For $\scG$ a crossed module of Lie groups $(\sfH\xrightarrow{~\partial}\sfG,\acton)$, this functor consists of two maps, $g:Y^{[2]}\coloneqq Y\times_MY\rightarrow\sfG$ and $h:Y^{[3]}\coloneqq Y\times_MY\times_MY\rightarrow\sfH$ such that 
            \begin{equation}
                \begin{aligned}
                    h_{y_1y_3y_4}h_{y_1y_2y_3}\ &=\ h_{y_1y_2y_4}(g_{y_1y_2}\acton h_{y_2y_3y_4})~,
                    \\
                    g_{y_1y_3}\ &=\ \partial(h_{y_1y_2y_3})g_{y_1y_2}g_{y_2y_3}
                \end{aligned}
            \end{equation}
            for all $(y_1,\ldots,y_4)\in Y^{[4]}$ and $(y_1,y_2,y_3)\in Y^{[3]}$, respectively. Natural isomorphisms between two such functors $\Gamma,\tilde\Gamma:\check\scC(\sigma)\rightarrow\sfB\scG$ are given by maps $a:Y\rightarrow\sfG$ and $b:Y^{[2]}\rightarrow\sfH$ such that
            \begin{equation}
                \begin{aligned}
                    \tilde h_{y_1y_2y_3}\ &=\ a_{y_1}^{-1}\acton(b_{y_1y_3}h_{y_1y_2y_3}(g_{y_1y_2}\acton b_{y_2y_3}^{-1})b_{y_1y_2}^{-1})~,
                    \\
                    \tilde g_{y_1y_2}\ &=\ a_{y_1}^{-1}\partial(b_{y_1y_2})g_{y_1y_2}a_{y_2}
                \end{aligned}
            \end{equation}
            for all $(y_1,y_2,y_3)\in Y^{[3]}$ and $(y_1,y_2)\in Y^{[2]}$, respectively. Finally, modifications between two such natural transformations are given by maps $c:Y\rightarrow\sfH$ such that
            \begin{equation}
                \begin{aligned}
                    \tilde b_{y_1y_2}\ &=\ c_{y_1}b_{y_1y_2}(g_{y_1y_2}\acton c_{y_2}^{-1})~,
                    \\
                    \tilde a_y\ &=\ \partial(c_y)a_y
                \end{aligned}
            \end{equation}
            for all $(y_1,y_2)\in Y^{[2]}$ and $y\in Y$, respectively.
            
            Principal $\scG$-bundles are classified in terms of the non-Abelian cohomologies defined by the above \v Cech cohomology~\cite{Giraud:1971}, see also~\cite{Wockel:2008aa,Nikolaus:2011ag} for a proof. Here, the \v Cech cocycles are identified with the functors $\Gamma:\check\scC(\sigma)\rightarrow\sfB\scG$ and the \v Cech cochains are identified with natural isomorphisms between these.
        \end{example}
        
        \begin{remark}
            Principal $2$-bundles over a manifold $M$ can often be related to principal bundles over the loop space $LM$ of $M$. In particular, multiplicative Abelian gerbes can be \uline{transgressed} to circle bundles over loop space~\cite{Waldorf:0911.3212,Waldorf:2010aa}. Similarly, principal $2$-bundles with structure $2$-group $\sfString(n)$ from \cref{ex:string_2_group} are related to $\sfSpin(n)$-bundles over the loop space of the base space~\cite{mclaughlin1992orientation}. For a general discussion of transgression for principal $2$-bundles, see~\cite{Nikolaus:1112.4702}.
        \end{remark}
        
        \subsection{Connections on higher principal bundles}
        
        Connections on higher principal bundles are globalisations or integrations of corresponding local $L_\infty$-algebra valued connections discussed in \cref{ssec:BRST-complex}. The abstract construction of this integration procedure is found in~\cite{Fiorenza:2010mh}, see also~\cite{Fiorenza:2012tb}. Concrete formulas that are suitable for explicit computations for adjusted connections on principal $\sfString(\sfG)$-bundles were given in~\cite{Rist:2022hci}. 
        
        The first papers to discuss connections on non-Abelian gerbes are~\cite{Breen:math0106083,Aschieri:2003mw}, which present a notion of connection that contains additional data compared to that of the $L_\infty$-algebraic description. A corresponding description of non-Abelian $2$-gerbes is found in~\cite{Jurco:2009px}. For related early work on $\sfG$-gerbes, see~\cite{Laurent-Gengoux:2005wxa}. Underlying connections is the notion of higher parallel transport and higher holonomy, see~\cite{Gawedzki:2002se,Schreiber:2005mi,Schreiber:2008aa,Fuchs:2009dms,Martins:2011:3309,Schreiber:1303.4663,Wang:1512.08680,Waldorf:1704.08542} for details. For further literature on connections, see also~\cite{Waldorf:1608.00401,Zucchini:2019pbv,Zucchini:2019rpp}, which give a total space perspective on higher connections. For a detailed description of higher bundles with connections from the simplicial perspective see~\cite{Jurco:2005qj}.
        
        Below, we review the review the picture of~\cite{Rist:2022hci}, which gives an explicit form of the cocycles for adjusted connections on principal $2$-bundles with strict structure $2$-group; see also~\cite{Tellez-Dominguez:2023wwr}, which presents adjusted connections from a different perspective than~\cite{Rist:2022hci}.
        
        \paragraph{Adjustment for strict $2$-groups.}
        In the picture of~\cite{Rist:2022hci}, an \uline{adjustment datum} $\kappa$ for a crossed module of Lie groups $(\sfH\xrightarrow{~\partial~}\sfG,\acton)$ is a map 
        \begin{subequations}
            \begin{equation}
                \kappa\,:\,\sfG\times\frg\ \rightarrow\ \frh
            \end{equation}
            with $\frg$ and $\frh$ the Lie algebras of $\sfG$ and $\sfH$, respectively, such that
            \begin{equation}\label{eq:adjustment_condition}
                \begin{aligned}
                    \kappa(\partial(h),V)\ &=\ h(V\acton h^{-1})~,
                    \\
                    \kappa(g_2g_1,V)\ &=\ g_2\acton\kappa(g_1,V)+\kappa\big(g_2,g_1Vg^{-1}_1-\partial(\kappa(g_1,V))\big)
                \end{aligned}
            \end{equation}
        \end{subequations}        
        for all $g_{1,2}\in\sfG$, $h\in\sfH$, and $V\in\frg$.
        
        \paragraph{Adjustment differential refinement of principal $2$-bundles.}
        The adjusted differential refinement of a \v Cech cocycle describing a principal $(\sfH\xrightarrow{~\partial~}\sfG,\acton)$-bundle is then given by the data from \cref{exa:2Bundle} and a tuple $(A,B,\Lambda)$ with 
        \begin{subequations}\label{eq:adjustedCocycleConditions}
            \begin{equation}
                a\ =\ A+B+\Lambda\ \in\ \Omega^1(Y)\otimes\frg\,\oplus\,\Omega^2(Y)\otimes\frh\,\oplus\,\Omega^1(Y^{[2]})\otimes\frh
            \end{equation}
            such that
            \begin{equation}\label{eq:adjustedCocycleConditionsB}
                \begin{aligned}
                    \Lambda_{y_1y_3}\ &=\ \Lambda_{y_2y_3}+g_{y_2y_3}^{-1}\acton \Lambda_{y_1y_2}-g_{y_1y_3}^{-1}\acton(h_{y_1y_2y_3}\nabla_{y_1} h_{y_1y_2y_3}^{-1})~,
                    \\
                    A_{y_2}\ &=\ g^{-1}_{y_1y_2}A_{y_1}g_{y_1y_2}+g^{-1}_{y_1y_2}\rmd g_{y_1y_2}-\partial(\Lambda_{y_1y_2})~,
                    \\
                    B_{y_2}\ &=\ g^{-1}_{y_1y_2}\acton B_{y_1}+\rmd\Lambda_{y_1y_2}+ A_{y_2}\acton\Lambda_{y_1y_2}+\tfrac12[\Lambda_{y_1y_2},\Lambda_{y_1y_2}]
                    \\
                    &\hspace{1cm}-\kappa\big(g_{y_1y_2}^{-1},\rmd A_{y_1}+\tfrac12[A_{y_1},A_{y_1}]+\partial(B_{y_1})\big)\,,
                \end{aligned}
            \end{equation}
        \end{subequations}
        for all appropriate $(y_1,y_2,\ldots)\in Y^{[n]}$; here $\nabla_y\coloneqq\rmd+A_y\,\acton$. Bundle isomorphisms $(a,b)$ from \cref{exa:2Bundle} between differentially refined cocycles $(g,h,A,B,\Lambda)$ and $(\tilde g,\tilde h,\tilde A,\tilde B,\tilde\Lambda)$ are differentially refined by an additional map $\lambda\in\Omega^1(Y)\otimes\frh$ such that 
        \begin{equation}\label{eq:adjustedCoboundaryConditionsB}
            \begin{aligned}
                \tilde\Lambda_{y_1y_2}\ &=\ a^{-1}_{y_2}\acton\Lambda_{y_1y_2}+\lambda_{y_2}-\tilde{g}^{-1}_{y_1y_2}\acton\lambda_{y_1}+(a_{y_2}^{-1}g_{y_1y_2}^{-1})\acton(b_{y_1y_2}^{-1}\nabla_{y_1}b_{y_1y_2})~,                    
                \\
                \tilde A_y\ &=\ a_y^{-1}A_ya_y+a_y^{-1}\rmd a_y-\partial(\lambda_y)~,
                \\
                \tilde B_y\ &=\ a_y^{-1}\acton B_y+\rmd\lambda_y+\tilde A_y\acton\lambda_y+\tfrac12[\lambda_y,\lambda_y]-\kappa\big(a_y^{-1},F_y\big)
            \end{aligned}
        \end{equation}
        for all $y\in Y$ and $(y_1,y_2)\in Y^{[2]}$. Finally, higher gauge transformations $c$ relate differential refinements $\lambda$ of the bundle isomorphism according to
        \begin{equation}
            \tilde\lambda_y\ =\ \lambda_y+a_y^{-1}\acton(c_y^{-1}\nabla_yc_y)
        \end{equation}
        for all $y\in Y$.
        
        The local curvature forms of the local connections $A$ and $B$ are the same as in the local situation
        \begin{equation}\label{eq:adjustedCurvatures}
            \begin{aligned}
                F_y\ &\coloneqq\ \rmd A_y+\tfrac12[A_y,A_y]+\partial(B_y)\ \in\ \Omega^2(Y)\otimes\frg~,
                \\
                H_y\ &\coloneqq\ \rmd B_y+A_y\acton B_y-\kappa(A_y,F_y)\ \in\ \Omega^3(Y)\otimes\frh~,
            \end{aligned}
        \end{equation}
        and the Bianchi identities read as
        \begin{equation}\label{eq:adjustedBianchiIdentities}
            \nabla_y F_y\ =\ \partial(H_y+\kappa(A_y,F_y))
            \eand
            \nabla_y H_y\ =\ \kappa(A_y,\partial(H_y))-\kappa(F_y,F_y)
        \end{equation}
        for all $y\in Y$.
        
        Note that the adjustment fixes the additional data arising in the connections on non-Abelian gerbes of~\cite{Breen:math0106083,Aschieri:2003mw} in terms of other cocycle data.
        
        \begin{example}
            Particularly interesting examples are principal $2$-bundles with structure $2$-group the $2$-group model of the string group of~\cref{ex:string_2_group} because these correspond to string structures, higher versions of spin structures, cf.~\cite{Waldorf:2009uf} and~\cite{Waldorf:2023zmt}. A geometrically very natural such bundle arises from the lift of the principal $\sfSpin(4)$-bundle $\sfSpin(5)\rightarrow\sfSpin(5)/\sfSpin(4)\cong S^4$, as pointed out in~\cite{Roberts:2022wwl}. This is a doubling of the fundamental instanton given by the quaternionic Hopf fibration to a fundamental instanton--anti-instanton pair, which possesses a topologically unique lift to a $\sfString(4)$-bundle. An explicit cocycle description for this principal $2$-bundle, including the differential refinement in form of an adjusted connection has been given in~\cite{Rist:2022hci}. See also~\cite{Demessie:2016ieh} for the unadjusted case, as well as~\cite{Sheng:1701.00959} for the construction of an Atiyah algebroid perspective.
        \end{example}
        
        Forcing $\kappa=0$ (which is not an adjustment datum!), one recovers unadjusted connections\footnote{as well as gauge transformations, curvatures, and Bianchi identities} used in earlier literature, cf.~e.g.~\cite{Schreiber:2008aa}, where a form of parallel transport is discussed; see also~\cite{Kim:2019owc} for details on an adjusted form of parallel transport. For other work on parallel transport, see also~\cite{Schreiber:1303.4663,Soncini:2014zra,Waldorf:1704.08542}. 
        
        \section{Dynamics: applications and examples}\label{sec:dynamics}
        
        \subsection{Higher Chern--Simons theories}
        
        \paragraph{Higher Chern--Simons theory.}
        Let $M$ be a $n$-dimensional closed oriented manifold and $\frL=\bigoplus_{k=-n+3}^0\frL_k$ be a cyclic $(n-2)$-term $L_\infty$-algebra with inner product $\inner{-}{-}_\frL$. There is a natural tensor product $L_\infty$-algebra $\Omega^\bullet(M)\otimes\frL$ of the differential graded commutative algebra given by the de~Rham complex $\Omega^\bullet(M)$ and $\frL$,
        \begin{subequations}\label{eq:HCSLInfinity}
            \begin{equation}
                \hat\frL\ \coloneqq\ \Omega^\bullet(M)\otimes\frL\ \coloneqq\ \bigoplus_{k\in\IZ}\hat\frL_k
                \ewith
                \hat \frL_k\ \coloneqq\ \sum_{\substack{i\geq0\\i+j=k}}\Omega^k(M)\,\otimes\,\frL_j~,
            \end{equation}
            \begin{equation}
                \begin{aligned}
                    \hat\mu_1(\alpha_1\otimes\ell_1)\ &\coloneqq\ \rmd\alpha_1\otimes\ell_1+(-1)^{|\alpha_1|_{\Omega^\bullet(M)}}\alpha_1\otimes\mu_1(\ell_1)~,
                    \\[5pt]
                    \hat\mu_i(\alpha_1\otimes\ell_1,\ldots,\alpha_i\otimes\ell_i)\ &\coloneqq\ (-1)^{i\sum_{j=1}^i|\alpha_j|_{\Omega^\bullet(M)}+\sum_{j=0}^{i-2}|\alpha_{i-j}|_{\Omega^\bullet(M)}\sum_{k=1}^{i-j-1}|\ell_k|_\frL}
                    \\[-5pt]
                    &\kern2cm\times(\alpha_1\wedge\ldots\wedge\alpha_i)\otimes\mu_i(\ell_1,\ldots,\ell_i)
                \end{aligned}
            \end{equation}
        \end{subequations}
        for all $\alpha_{1,\ldots,i}\in\Omega^\bullet(M)$ and $\ell_{1,\ldots,i}\in\frL$. This tensor product comes with a cyclic inner product 
        \begin{equation}
            \inner{\alpha_1\otimes\ell_1}{\alpha_2\otimes\ell_2}_{\hat\frL}\ \coloneqq\ (-1)^{|\ell_1|_\frL|\alpha_2|_{\Omega^\bullet(M)}}\int\alpha_1\wedge\alpha_2\,\inner{\ell_1}{\ell_2}_\frL
        \end{equation}
        of degree~$-3$. Consequently, for $a\in\hat\frL$ of degree~$1$, we can consider the action
        \begin{equation}\label{eq:hCStheory}
            S\ \coloneqq\ \sum_i\frac{1}{(i+1)!}\inner{a}{\hat\mu_i(a,\ldots,a)}_{\hat\frL}~.
        \end{equation}
        For $n=3$, we have $a=A\in\Omega^1(M)\otimes\frL_0$, and this action reduces to that of standard Chern--Simons theory. Likewise, for $n=4$, we have $a=A+B\in\Omega^1(M)\otimes\frL_0\,\oplus\,\Omega^2(M)\otimes\frL_{-1}$, and this action becomes that of higher Chern--Simons theory with a $2$-group as gauge group. 
        
        This four-dimensional Chern--Simons theory is a generalised form of a $BF$-theory, which has been studied e.g.~in~\cite{Soncini:2014ara}, see also~\cite{Zucchini:2015ohw} and~\cite{Zucchini:2021bnn}. It should not be confused with the four-dimensional Chern--Simons theory discussed in~\cite{Costello:2013zra}. Just as Chern--Simons theory can be used to compute knot invariants, higher Chern--Simons theories may produce higher holonomy invariants~\cite{Zucchini:2015wba,Zucchini:2015xba}.
        
        \paragraph{Higher holomorphic Chern--Simons theory.} 
        In the above construction, we can replace the de~Rham complex by the Dolbeault complex $(\Omega^{0,\bullet}(M),\bar\partial)$ of a closed complex manifold $M$, and the tensor product $L_\infty$-algebra~\eqref{eq:HCSLInfinity} generalises in the evident way to
        \begin{equation}
            \hat\frL\ \coloneqq\ \Omega^{0,\bullet}(M)\otimes\frL~.
        \end{equation}
        If the complex $n$-dimensional manifold $M$ is Calabi--Yau, i.e.~it admits a globally defined holomorphic volume form $\Omega\in\Omega^{0,n}(M)$, then there is an evident cyclic structure 
        \begin{equation}
            \inner{\alpha_1\otimes\ell_1}{\alpha_2\otimes\ell_2}_{\hat\frL}\ \coloneqq\ (-1)^{|\ell_1|_\frL|\alpha_2|_{\Omega^{0,\bullet}(M)}}\int\Omega\wedge\alpha_1\wedge\alpha_2\,\inner{\ell_1}{\ell_2}_\frL~.
        \end{equation}
        Using this inner product, we again obtain an action of the form~\eqref{eq:hCStheory}. For $n=3$, this becomes standard holomorphic Chern--Simons theory, and for $n>3$ holomorphic higher Chern--Simons theory, respectively. An application of this theory is given in~\cref{ssec:twistors}.
        
        \subsection{Alexandrov--Kontsevich--Schwarz--Zaboronsky sigma models}\label{ssec:AKSZ}
        
        For a review of the Alexandrov--Kontsevich--Schwarz--Zaboronsky (AKSZ) construction~\cite{Alexandrov:1995kv}, see e.g.~\cite{Cattaneo:2010re,Qiu:1105.2680,Bouwknegt:2011vn,Ikeda:2012pv,Mnev:2019aa} as well as~\cite{Park:2000au,Hofman:2002rv,Pulmann:2019vrw} for some applications.
        
        \paragraph{Setting.}
        In the seminal AKSZ construction, one starts with a \uline{symplectic N$Q$-manifold} $(M,Q_M,\omega_M)$ of degree $n$, i.e.~a non-negatively graded smooth manifold $M$ equipped with a nilquadratic vector field $Q_M$ of degree~$1$, also known as a \uline{homological vector field}, together with a symplectic form $\omega_M$ of degree $n$ such that $Q_M$ is symplectic, i.e.~$\omega_M$ is $Q_M$-invariant. Furthermore, let $\Sigma$ be an oriented closed $(n+1)$-dimensional smooth manifold. Functions on the shifted tangent bundle $T[1]\Sigma$ are canonically identified with differential forms on $\Sigma$, hence the de~Rham differential $\rmd_\Sigma$ on $\Sigma$ defines a homological vector field $Q_{T[1]\Sigma}$ on $T[1]\Sigma$. Let $\mu$ be a non-degenerate $Q_{T[1]\Sigma}$-invariant measure of degree $-n-1$ on $T[1]\Sigma$. Then, the space of smooth maps
        \begin{equation}
            \caM\ \coloneqq\ \scC^\infty(T[1]\Sigma,M)
        \end{equation}
        is a Batalin--Vilkovisky (BV) manifold, i.e.~an infinite-dimensional graded smooth manifold equipped with the degree~$-1$ symplectic form
        \begin{subequations}\label{eq:AKSZSymplecticForm}
            \begin{equation}
                \varpi_\caM\ \coloneqq\ \int_{T[1]\Sigma}\mu\,\sfev^*\omega_M~,
            \end{equation}
            where 
            \begin{equation}
                \begin{aligned}
                    \sfev\,:\,\caM\times T[1]\Sigma\ &\rightarrow\ M~,
                    \\
                    (f,x)\ &\mapsto\ f(x)
                \end{aligned}
            \end{equation}
        \end{subequations}
        for all $(f,x)\in\caM\times T[1]\Sigma$. Furthermore, diffeomorphisms on $T[1]\Sigma$ and $M$ induce actions on $\caM$ by pre- or post-composition, respectively. Therefore, the two homological vector fields $Q_{T[1]\Sigma}$ and $Q_M$ induce homological vector fields $\caQ_{T[1]\Sigma}$ and $\caQ_M$ on $\caM$, respectively. We can then choose any (constant) linear combination of these vector fields to form a homological vector field $\caQ_\caM$ on $\caM$ so that $\varpi_\caM$ is $\caQ_\caM$-invariant~\cite{Cattaneo:2001ys,Roytenberg:2006qz}. In the following, we shall take 
        \begin{equation}\label{eq:AKSZHomologicalVectorField}
            \caQ_\caM\ \coloneqq\ \caQ_{T[1]\Sigma}+\caQ_M~.
        \end{equation}
        The degree~$1$ Poisson bracket associated with $\varpi_\caM$, also known as the \uline{anti-bracket} or \uline{BV bracket}, is denoted by $\{-,-\}_\caM$.
        
        \paragraph{AKSZ construction.}
        In the AKSZ construction, the manifolds $\Sigma$ and $M$ are the respective \uline{source} and \uline{target manifolds} for a class of sigma models given by a classical BV action in the superfield formalism. The construction proceeds as follows. Associated with $Q_M$, we have the corresponding degree $n+1$ Hamiltonian $H_M\coloneqq\frac{1}{n+1}Q_M\intprod\Upsilon_M\intprod\omega_M$, for $n\neq -1$ and $\Upsilon_M$ the Euler vector field on $M$. It satisfies $\{H_M,H_M\}_M=0$ with $\{-,-\}_M$ the degree $-n$ Poisson bracket on $M$ corresponding to $\omega_M$. Moreover, if $\theta_M\in\Omega^1(M)$ is an arbitrary symplectic potential $1$-form, $\rmd_M\theta_M=\omega_M$, for example, $\theta_M=\frac{1}{n}\Upsilon_M\intprod\omega_M$, we can compute the Hamiltonian $S_\caM$ corresponding to the homological vector field $Q_\caM$ in~\eqref{eq:AKSZSymplecticForm} for the symplectic form $\varpi_\caM$ in~\eqref{eq:AKSZSymplecticForm}. In turn, $S_\caM$ is nothing but the AKSZ model action. In particular, for $f\in\caM$, we arrive at 
        \begin{equation}\label{eq:AKSZAction}
            S_\caM[f]\ \coloneqq\ \caQ_{T[1]\Sigma}\intprod\int_{T[1]\Sigma}\mu\,\sfev^*_{(f,x)}\theta_M+\int_{T[1]\Sigma}\mu\,\sfev^*_{(f,x)}H_M~.
        \end{equation}
        Upon integrating out the odd coordinates on $T[1]\Sigma$, we obtain the action as an integral over the source $\Sigma$. By construction, it solves the \uline{classical master equation},
        \begin{equation}
            \{S_\caM,S_\caM\}_\caM\ =\ 0~.
        \end{equation}
        Classical solutions, i.e~the critical points of the action $S_\caM$, are maps $f:T[1]\Sigma\rightarrow M$ intertwining between $\caQ_{T[1]\Sigma}$ and $\caQ_M$, i.e.~they are $\caQ_\caM$-preserving. Let us now discuss some examples.
        
        \begin{example}
            The \uline{Poisson sigma model} model is obtained as follows. Let $N$ be a smooth manifold and take the target $M\coloneqq T^*[1]N$ with its canonical symplectic from $\omega_M$ of degree~$1$. Correspondingly, take a two-dimensional source $\Sigma$ with local coordinates $\sigma^\mu$. We denote the degree~$1$ fibre coordinates of $T[1]\Sigma$ by $\theta^\mu$. Hence, the homological vector field $Q_{T[1]\Sigma}$ on $T[1]\Sigma$ is given by $Q_{T[1]\Sigma}=\theta^\mu\parder{\sigma^\mu}$. Furthermore, we denote the degree~$0$ local coordinates on $N$ as $x^i$ and the degree~$1$ fibre coordinates on $T^*[1]N$ by $\chi_i$. The homological vector field $Q_M$ is necessarily determined by a Poisson $2$-vector field on $N$, 
            \begin{equation}
                \pi_M\ \coloneqq\ \frac{1}{2}\pi^{ij}(x)\parder{x^i}\wedge\parder{x^j}
            \end{equation}
            with vanishing Schouten bracket.\footnote{In fact, isomorphism classes of symplectic N$Q$-manifolds of degree~$1$ are in one-to-one correspondence with isomorphism classes of Poisson manifolds~\cite{Severa:2001aa}.} Explicitly,
            \begin{equation}
                Q_M\ =\ \pi^{ij}(x)\chi_j\parder{x^i}+\frac12\parder[\pi^{ij}(x)]{x^k}\chi_i\chi_j\parder{\chi_k}~.
            \end{equation}
            The corresponding Hamiltonian is given by
            \begin{equation}
                H_M\ =\ \tfrac12\pi^{ij}(x)\chi_i\chi_j~.
            \end{equation}
            We take the symplectic potential $\theta_M=\chi_i\rmd_Mx^i$, and we use the notation $\bm x^i$, ${\bm\chi}_i$ for the pull-backs under $f$ of $x^i$ and $\chi_i$, respectively. The AKSZ action~\eqref{eq:AKSZAction} then becomes the action of the Poisson sigma model of~\cite{Ikeda:1993fh,Schaller:1994es}, see also~\cite{Cattaneo:2001ys},
            \begin{equation}
                S_\caM[\bmx,{\bm\chi}]\ =\ \int_{T[1]\Sigma}\mu\,\Big\{{\bm\chi}_iQ_{T[1]\Sigma}\bmx^i+\tfrac12\pi^{ij}(\bmx){\bm\chi}_i{\bm\chi}_j\Big\}\,.
            \end{equation}
            Solutions to the classical field equations are Lie algebroid maps $T[1]\Sigma\rightarrow M$, cf.~\cite{Severa:2001aa}.
        \end{example}
        
        \begin{example}
            To construct the \uline{Courant sigma model}, the starting point is the observation made in~\cite{Severa:2001aa,Roytenberg:0203110} that isomorphism classes of symplectic N$Q$-manifolds of degree~$2$ are in one-to-one correspondence with isomorphism classes of Courant algebroids. Let $E\rightarrow X$ be a Courant algebroid. We denote the local Darboux coordinates on the corresponding symplectic N$Q$-manifold $M$ of degrees~$0$, $1$, and $2$ by $x^i$, $e^a$, and $p_i$, respectively. The indices $i$ range from~$1$ to the dimension of the base $X$ and indices $a$ from $1$~to the dimension of the fibre. Correspondingly, we have the local structure functions of $E$. We use $\ttg_{ab}\in\IR$ for the pairing, $\tta^i_a\in\scC^\infty(X)$ for the anchor, and $\ttc_{abc}\in\scC^\infty(X)$ for the Dorfman bracket. We then take 
            \begin{equation}
                \theta_M\ \coloneqq\ p_i\rmd_Mx^i+\tfrac12\ttg_{ab}e^a\rmd_Me^b
                \eand
                H_M\ \coloneqq\ -\tta^i_a(x)p_ie^a+\tfrac{1}{3!}\ttc_{abc}(x)e^ae^be^c~,
            \end{equation}
            and the resulting in the AKSZ action~\eqref{eq:AKSZAction} becomes the action of the Courant sigma model~\cite{Ikeda:2002wh,Roytenberg:2006qz}
            \begin{equation}\label{eq:CourantSigmaModelAction}
                S_\caM[\bmx,\bme,\bmp]\ =\ \int_{T[1]\Sigma}\mu\,\Big\{\bmp_iQ_{T[1]\Sigma}\bmx^i+\tfrac12\ttg_{ab}\bme^aQ_{T[1]\Sigma}\bme^b-\tta^i_a(\bmx)\bmp_i\bme^a+\tfrac{1}{3!}\ttc_{abc}(\bmx)\bme^a\bme^b\bme^c\Big\}\,,
            \end{equation}
            where, as before, $Q_{T[1]\Sigma}=\theta^\mu\parder{\sigma^\mu}$ with $\sigma^\mu$ and $\theta^\mu$ the local coordinates of degrees~$0$ and $1$ on $T[1]\Sigma$, and we have also denoted the pull-backs of $x^i$, $e^a$, and $p_i$ under $f$ by $\bmx^i$, $\bme^a$, and $\bmp_i$, respectively. Note that according to the AKSZ construction, the source $\Sigma$ is three-dimensional.
        \end{example}
        
        \begin{example}\label{ex:cs_theory}
            \uline{Chern--Simons theory} is obtained as a special case of the Courant sigma model when the Courant algebroid $E$ is reduced to a metric Lie algebra $(\frg,\inner{-}{-})$. Consequently, are no $x^i$ and $p_i$ coordinates, and $M\coloneqq\frg[1]$. The homological vector field $Q_M$ is induced by the Chevalley--Eilenberg differential on $\frg$ and the invariant pairing induces on $\frg[1]$ a degree~$2$, $Q_M$-invariant symplectic form. In this case, $\caM\cong\Omega^\bullet(M)\otimes\frg[1]$ and the corresponding AKSZ action~\eqref{eq:AKSZAction} follows from~\eqref{eq:CourantSigmaModelAction} after dropping the $\bmx^i$ and $\bmp_i$ coordinates and integrating over the odd coordinates in $T[1]\Sigma$,
            \begin{equation}
                S_\caM[A]\ =\ \int_\Sigma\Big\{\tfrac12\inner{A}{\rmd_\Sigma A}+\tfrac1{3!}\inner{[A,A]}{A}\Big\}\,.
            \end{equation}
        \end{example}
        
        \paragraph{Relation to higher gauge theory.}
        In the Courant sigma model, the structure function $\ttc_{abc}(x)$ usually is taken to be a closed $3$-form, hence it possesses a local gauge potential $2$-form. Note, however, that there is, strictly speaking no requirement for $\ttc\in H^3(M,\IZ)$, so it is not necessarily the curvature of an Abelian gerbe. Also, \cref{ex:cs_theory} can be generalised to higher Chern--Simons theories in an evident manner. Finally, see also~\cite{Zucchini:2011aa,Zucchini:2017nax,Hyungrok:2018aa,Pulmann:2019vrw} for other formulations of higher gauge theory in the context of the AKSZ construction.

        \subsection{Higher gauge theories on twistor spaces}\label{ssec:twistors}
        
        Twistor theory originated from Penrose's desire to combine the complex geometry underlying quantum mechanics with the light cones of relativity theory. From a different perspective, recall that many gauge field equations can be regarded as flatness constraints on a gauge potential along certain subspaces. A twistor space is now the moduli space of such subspaces, and by the Penrose--Ward transform~\cite{Ward:1977ta,Penrose:1985jw,Penrose:1986ca}, holomorphic bundles and their gauge orbits over twistor space correspond to solutions to field equations and their gauge orbits. Twistor spaces manifest symmetries of the underlying field theories, and in particular supersymmetric extensions are easily constructed. For reviews, see e.g.~\cite{Popov:2004rb,Wolf:2010av}.
        
        For Abelian gerbes over $\IC^6$ with self-dual (complexified) $3$-form curvature, this was first studied in~\cite{Hughston:1979TN,Hughston:1987aa,Hughston:1988nz,0198535651,Berkovits:2004bw}, see also~\cite{Saemann:2011nb,Mason:2011nw,Mason:2012va}, and the relevant twistor space $P^6$ is a quadric inside the decompactification of $\IC P^7$ given by the total space of the rank-4 complex vector bundle $\IC^4\otimes\caO(1)\rightarrow\IC P^3$.
        
        The generalisation to the non-Abelian case, employing holomorphic versions of various forms of higher principal bundles over twistor space, was studied in~\cite{Saemann:2012uq,Saemann:2013pca,Jurco:2014mva,Jurco:2016qwv}. In the simplest case, one considers a principal $2$-bundle with strict structure $2$-group, described by holomorphic \v Cech cocycles on $P^6$, which is topologically trivial and holomorphically trivial when restricted to sections of $P^6$. The appropriate generalisation of the Penrose--Ward transform maps such a bundle to a connection of a topologically trivial principal $2$-bundle over $\IR^{1,5}$, which comes with a connection that has a (non-Abelian) self-dual $3$-form curvature. There is an evident supersymmetric extension of $P^6$, which leads to the field content of a $\caN=(2,0)$ tensor multiplet on $\IR^{1,5}$. The connection, however, suffers from being fake-flat, a seemingly unavoidable problem in the higher Penrose--Ward transform for six dimensions. An iterative construction which is less elegant, however, is possible~\cite{Rist:2022hci}.
        
        Another application of higher gauge theory in this context was given in~\cite{Samann:2017dah}. Recall that the definition of a higher holomorphic Chern--Simons action principle requires the underlying manifold to be Calabi--Yau. A generalisation was given in~\cite{Witten:2003nn}, where a supersymmetric extension of holomorphic Chern--Simons theory was considered. This works well for the twistor space describing of maximally supersymmetric self-dual Yang--Mills theory. For the superambitwistor space $L^{5|6}$, which describes solutions to the full $\caN=3$ supersymmetric Yang--Mills equations, this is however not sufficient because the ambitwistor space is (bosonic) five-dimensional. There is, however, a natural higher holomorphic Chern--Simons theory on this space using a higher principal bundle with structure $3$-group. With the appropriate choice of this $3$-group (or after taking equivalences), the solutions and gauge orbits to this higher holomorphic Chern--Simons theory are equivalent to the solutions and gauge orbits of $\caN=3$ supersymmetric Yang--Mills theory~\cite{Samann:2017dah}.
        
        \subsection{Higher Stueckelberg model}\label{ssec:Stueckelberg}
        
        Let $M$ be a closed oriented (semi-)Riemannian manifold and consider the ordinary action
        \begin{equation}
            S\ \coloneqq\ \tfrac12\int\big\{\rmd C\wedge\star\rmd C+m^2C\wedge\star C\big\}
        \end{equation}
        for a massive $p$-form gauge field $C\in\Omega^p(M)$. This action does not enjoy a gauge symmetry and, as such, it is not a (higher) gauge theory. However, it is often convenient to introduce (higher) gauge symmetry into the system, resulting in the \uline{Stueckelberg action}\footnote{originally due to Stueckelberg~\cite{Stueckelberg:1938aa}; the $p$-form variant is due to~\cite{Bizdadea:1996np}}
        \begin{equation}
            S_\mathrm{St}\ \coloneqq\ \tfrac12\int\big\{\rmd C\wedge\star\rmd C+m^2(\rmd B+C)\wedge\star(\rmd B+C)\big\}
        \end{equation}
        which, in addition to the original $p$-form field $C$, also contains the \uline{Stueckelberg $(p-1)$-form field} $B\in\Omega^{p-1}(M)$. The action $S_\mathrm{St}$ enjoys the gauge symmetry
        \begin{equation}
            C\ \mapsto\ C+\rmd\Lambda
            \eand
            B\ \mapsto\ B-\Lambda+\rmd\lambda
        \end{equation}
        for all $\Lambda\in\Omega^{p-1}(M)$ and $\lambda\in\Omega^{p-2}(M)$. As such, this is a gauge theory with (higher) gauge $L_\infty$-algebra
        \begin{equation}
            \frL\ \coloneqq\ (\cdots\xrightarrow{~~~}*\xrightarrow{~~~}\IR[p-1]\xrightarrow{~\sfid~}\IR[p-2]\xrightarrow{~~~}*\xrightarrow{~~~}\cdots)~,
        \end{equation}
        and evident higher gauge $p$-group. We note that $\frL$ is quasi-isomorphic to the trivial $L_\infty$-algebra, as they have the same cohomology and hence minimal model. At the group level, the $p$-group is then Morita equivalent to the trivial $p$-group. Yet, the Stueckelberg action $S_\mathrm{St}$ is physically equivalent to $S$, as is evident from making the gauge choice $B=0$.
        
        This is an example of a higher gauge theory whose gauge $\infty$-group is Morita-equivalent to a trivial one but which still contains non-trivial kinematic data non-trivial dynamics; it illustrates the fact that Morita equivalence of the gauge $\infty$-group is coarser than physical equivalence.
        
        \subsection{Supergravity and T-duality}\label{ssec:Tdualities}
        
        \paragraph{$B$-field and Abelian gerbes.}
        Just as a charged point-particle, tracing out a world-line $\tau_{\text{particle}}:\Sigma_1\rightarrow M$ through a closed oriented (semi-)Riemannian manifold $M$, couples to a $1$-form gauge potential $A$, 
        \begin{equation}
            \int\tau_{\text{particle}}^*A
        \end{equation}
        a string, tracing out  a world-sheet $\tau_{\text{string}}:\Sigma_2\rightarrow M$, couples to a $2$-form gauge potential $B$
        \begin{equation}
            \int\tau_{\text{string}}^*B~.
        \end{equation}
        This is the famous Kalb--Ramond $B$-field of string theory~\cite{Kalb:1974yc}. In the low-energy effective field theory limit, $B$ appears as a higher gauge field of supergravity. For $\sigma:Y\rightarrow M$ a cover of $M$, locally $B_y\in\Omega^2(Y)$ for all $y\in Y$ with $3$-form curvature given by
        \begin{equation}
            H_y\ =\ \rmd B_y~.
        \end{equation}
        In isolation, globally the $B$-field is a part of a connection on an Abelian principal $2$-bundle, or gerbe, discussed in~\cref{ssec:higher_principle_bundles}~\cite{Gawedzki:1987ak,Freed:1999vc}. If one considers supergravity in topologically non-trivial backgrounds, the  principal $2$-bundle picture is crucial. In particular, consistent string-coupling implies that locally integrating the $B$-field over the string world-sheet produces a $\sfU(1)$ holonomy, which implies that $B$ is globally a degree~$3$ Deligne cocycle, which is equivalent to an Abelian gerbe with connection, cf.~e.g.~\cite{Gawedzki:1987ak}.  
        
        \paragraph{$B$-field and general principal $2$-bundles.}
        However, viewing $B$ as a connection on an Abelian gerbe is generically insufficient.   For instance, on an orientifold background $B$ becomes a connection on a non-Abelian $\sfAut(\sfU(1))$-principal $2$-bundle~\cite{Schreiber:2005mi,Distler:2009ri,Deligne:1999qp}. Indeed, the potential need for non-Abelian generalisations was recognised early in the supergravity literature,  e.g.~see~\cite{Freedman:1980us,Freund:1981qw, Nepomechie:1982rb} for early attempts to construct  non-Abelian 2-form theories on  manifolds. 
        
        Moreover, the $B$-field will typically couple to other $p$-form fields in a variety of contexts that naturally arise in string theory and supergravity. In such cases, its characterisation as a connection requires more general higher gauge theories. Indeed, upon dimensionally reducing and gauging type II supergravity, the $B$-field will typically constitute merely a part of a higher gauge theory including $(p>2)$-form fields, as discussed in \cref{ssec:tensor_hierarchies}.  
        
        An important illustrative example is provided by Einstein--Maxwell/Yang--Mills supergravity theories, such as heterotic supergravity or type I supergravity coupled to  supersymmetric Yang--Mills multiplets. In such cases, it was long ago shown~\cite{Bergshoeff:1981um,Chapline:1982ww} that the Kalb--Ramond field-strength must be modified by a Chern--Simons $3$-form term deriving from the Yang--Mills $1$-form gauge potential $A$,
        \begin{equation}\label{eq:HandCSForm}
            H\ =\ \rmd B+\operatorname{cs}(A)
            \ewith
            \operatorname{cs}(A)\ \coloneqq\ \inner{A}{\rmd A}+\tfrac13\inner{A}{[A,A]}~.
        \end{equation}
        This promotes the $B$-field to a connection on a non-Abelian $2$-bundle, cf.~\cite{Baez:2005sn}. However, the Chern--Simons $3$-form does not arise directly from the standard Weil algebra associated to the corresponding $2$-term $L_\infty$ gauge algebra. Physically, the Yang--Mills fields are dynamical and do not satisfy the fake-flatness condition. This requires the use of an adjusted Weil algebra, as described in~\cref{ssec:local_adjustments}. 
        
        Note that anomaly cancellation via the Green--Schwarz mechanism requires the further addition of the Chern--Simons $3$-form $\operatorname{cs}(\omega)$ for the spin connection $\omega$.
        
        \paragraph{T-duality from principal $2$-bundles.} 
        In~\cite{Nikolaus:2018qop}, it was shown that geometric T-dualities, at the topological level, can be encoded in a principal $2$-bundle with gauge group $\sfTD_n$ as defined in \cref{ex:TDn}. A generic geometric string background, consisting of an Abelian gerbe on top of a principal torus bundle on top of a manifold can be described by a principal $2$-bundle with a particular structure group $\sfT\sfB_n^\text{F2}$. Such a principal $2$-bundle can be lifted to a principal $\sfTD_n$ bundle because the classifying spaces $\sfB\sfT\sfB_n^\text{F2}$ and $\sfB\sfTD_n$ are homotopy equivalent. On $\sfTD_n$, however, one has a manifest action of the T-duality group $\sfO(n,n,\IZ)$, which describes, among other things, the action of a T-duality. This picture was extended to affine torus bundles and further differentially refined in~\cite{Kim:2022opr,Kim:2023hqx}, see also~\cite{Kim:2023hqx} for a short review, with the crucial ingredient being an adjusted connection. The differential refinement also involves a generalisation of the structure $2$-group $\sfTD_n$ to a $\sfTD_n$-fibration over the Narain moduli space encoded in the $2$-groupoid $\scT\!\scD_n$. Altogether, a geometric T-duality, including the $B$-field as well as the dilaton field and a connection on the involved torus bundles (with the latter two encoding the Kaluza--Klein metric), can be regarded as a principal $\scT\!\scD_n$-bundle with adjusted connection.
        
        For further literature on T-duality from a higher perspective, see also~\cite{Fiorenza:2016oki,Fiorenza:2017jqx,Fiorenza:2018ekd,Sati:2018tvj}.
        
        \paragraph{Supergravity in $11$ dimensions.} 
        Similar considerations as in the above discussion should also arise in $11$-dimensional supergravity and therefore also in M-theory, but much less is known here. For some references, see e.g.~\cite{Aschieri:2004yz,Sati:2010ss,Sati:2010dc,Sati:2018tvj,Fiorenza:2018ekd,Sati:2021rsd}.
        
        \subsection{Tensor hierarchies}\label{ssec:tensor_hierarchies}
        
        The kinematical data in the previous example first appeared in equivalent form in the context of heterotic supergravity~\cite{Bergshoeff:1981um,Chapline:1982ww}. Gauged supergravities (reviewed in~\cite{Samtleben:2008pe,Weidner:2006rp,Trigiante:2016mnt}) generalise heterotic supergravities and feature series of $p$-form fields that are tightly constrained by representation-theoretic requirements, known as \uline{tensor hierarchies}~\cite{deWit:2005hv,deWit:2008ta}, which contain and generalise the Kalb--Ramond field (see \cref{ssec:Tdualities}) and Ramond--Ramond fields of ungauged supergravity.
        Being $p$-forms, the constraints arise from adjusted higher gauge symmetries in thin disguise~\cite{Borsten:2021ljb}. In addition to gauged supergravities, tensor hierarchies also appear in six-dimensional $\caN=(1,0)$ superconformal theories~\cite{Samtleben:2011fj,Samtleben:2012fb,Samtleben:2012mi} and more general Maxwell--Einstein-type theories~\cite{Bergshoeff:2009ph,Hartong:2009vc} with similar higher gauge symmetries.
        
        Briefly, in a gauged supergravity, one gauges a suitable subgroup $\sfG$ of the U-duality group $E_{k(k)}$, which is the split form of the $E$-series simple Lie groups, and the various $p$-form fields follow suitable representations of $\sfG$ determined by various constraints. This data combines into a differential graded Lie algebra. A shift-truncation construction~\cite{Borsten:2021ljb}, which is a refinement of the construction of~\cite{Getzler:1010.5859,Fiorenza:0601312}, turns this differential graded Lie algebra into an algebra over the operad $h\opLie$ in which the Lie bracket is only graded antisymmetric up to an alternator. The alternator now naturally produces the adjustment data for the higher gauge group (not to be confused with the ordinary group $\sfG$), as mentioned in \cref{ssec:local_adjustments}.
        
        \subsection{Superconformal field theories in six dimensions}\label{ssec:6dCFT}
        
        A special case of the tensor hierarchies gauge theory is that of a $(1,0)$-theory, which is superconformal at least at the classical level. The model was given in its first form in~\cite{Samtleben:2011fj}, with interpretations in terms of higher gauge theory found in~\cite{Palmer:2013pka,Lavau:2014iva}. The interpretation as an adjusted higher gauge theory was then developed in~\cite{Saemann:2017rjm,Saemann:2017zpd,Saemann:2019dsl,Rist:2020uaa}, with the last paper providing an improved mechanism for self-duality.\footnote{In this context, recall also the work of~\cite{Saemann:2012uq,Saemann:2013pca,Jurco:2014mva,Jurco:2016qwv}, in which six-dimensional superconformal field equations, albeit fake-flat ones, were derived on $\IR^{1,5|8}$ and $\IR^{1,5|16}$.}
        
        The model itself consists of a supersymmetric action on $\IR^{1,5}$ for a six-dimensional $(1,0)$-tensor multiplet with fields 
        \begin{subequations}\label{eq:1_0_theory}
            \begin{equation}
                \begin{gathered}
                    B\ \in\ \Omega^2(\IR^{1,5})\otimes(\IR_r\oplus \IR^*_s)~,
                    \\
                    \chi\ \in\ \Gamma(\IR^{1,5},\caS)\otimes (\IR_r\oplus \IR^*_s)~,~~~
                    \phi\ \in\ \scC^\infty(\IR^{1,5})\otimes(\IR_r\oplus\IR^*_s)~,
                \end{gathered}
            \end{equation}
            where $\caS$ denotes the bundle of Weyl spinors in six dimensions, a six-dimensional $(1,0)$-vector multiplet with fields taking values in a Lie algebra $\frg$,
            \begin{equation}
                \begin{gathered}
                    A\ \in\ \Omega^1(\IR^{1,5})\otimes(\frg_t\oplus\IR^*_p)~,
                    \\
                    \lambda\ \in\ \Gamma(\IR^{1,5},\caS)\otimes(\frg_t\oplus\IR^*_p)~,~~~
                    Y\ \in\ \scC^\infty(\IR^{1,5})\otimes\IR^3_{(ij)}\otimes(\frg_u\oplus\IR^*_p)~,
                \end{gathered}
            \end{equation}
            and a number of auxiliary fields
            \begin{equation}
                \begin{gathered}
                    C\ \in\ \Omega^3(\IR^{1,5})\otimes\frg_u^*\oplus\IR_q~,~~~
                    D\ \in\ \Omega^4(\IR^{1,5})\otimes\frg_v^*~,
                    \\
                    \beth_s\ \in\ \Omega^3_+(\IR^{1,5})\otimes\IR_s^*~,~~~
                    \daleth_t\ \in\ \Omega^2(\IR^{1,5})\otimes\frg_t~.
                \end{gathered}
            \end{equation}
            The subscripts in the above label the subspaces in a Lie $4$-algebra, given by a $4$-term $L_\infty$-algebra in which the various components of the fields take values. For example $A=A_t+A_p$ with $A_p\in\Omega^1(\IR^{1,5})\otimes\frg_t\oplus\IR^*_p$. The Lagrangian reads as~\cite{Rist:2020uaa}
            \begin{equation}\label{eq:1-0-Lagrangian}
                \begin{aligned}
                    \caL\ &\coloneqq\ -\rmd\phi_s\wedge\star\rmd\phi_r -\star4\bar\chi_s\slashed{\partial}\chi_r+\star4\bar\chi_s\inner{\slashed{F}_{\!t}}{\lambda_t}-\star8\bar\chi^i_s\inner{Y_{tij}}{\lambda^j_t}
                    \\  
                    &\hspace{1cm}+\phi_s\big(\inner{F_t}{\star F_t}-\star2\inner{Y_{tij}}{Y_t^{ij}}+\star4\inner{\bar\lambda_t}{\slashed{\nabla}\lambda_t}\big)
                    \\
                    &\hspace{1cm}+(\rmd B_s)^+\wedge\big(\rmd B_r+\operatorname{cs}(A_r)+\inner{\bar\lambda_t}{\gamma_{(3)}\lambda_t}\big)+(\rmd B_s)^-\wedge C_q^+   
                    \\
                    &\hspace{1cm}-\beth_s\wedge\big(\rmd B_r+\operatorname{cs}(A_r)-C_q+\inner{\bar\lambda_t}{\gamma_{(3)}\lambda_t}\big)
                    \\
                    &\hspace{1cm}+\daleth_t(\nabla C_u)+2B_s\wedge\inner{F_t}{\daleth_t}+\daleth_t(D_v)-2\phi_s\inner{\daleth_t}{\star F_t}
                    \\
                    &\hspace{1cm}-\star4\bar\chi_s\inner{\slashed{\daleth}_t}{\lambda_t}-B_s\wedge\inner{\daleth_t}{\daleth_t}+\phi_s\inner{\daleth_t}{\star\daleth_t}~.
                \end{aligned}
            \end{equation}
        \end{subequations}        
        Here, $F_t$ is the curvature of $A_t$, $\operatorname{cs}(-)$ is the Chern--Simons form appearing in~\eqref{eq:HandCSForm}, and $\pm$ refer to the projections onto the self-dual/anti-self-dual parts. Solutions to the equations of motion of this action describe a $(1,0)$-tensor multiplet with self-dual $3$-form curvature, as well as a generalised self-duality between the $2$- and $4$-form curvatures. These curvatures are the adjusted curvatures that can be constructed as for tensor hierarchies of gauged supergravity, cf.~\cref{ssec:tensor_hierarchies}. Self-duality is implemented purely with Lagrange multipliers, and the action can be supplemented by six-dimensional $(1,0)$-hypermultiplets~\cite{Samtleben:2012fb,Saemann:2017zpd,Rist:2020uaa}. The action restricts to both M2-brane models, as well as Yang--Mills theory in four dimensions~\cite{Saemann:2017zpd,Saemann:2017rjm}. Its BPS states, reduced to four dimensions, are non-Abelian versions of self-dual strings~\cite{Akyol:2012cq,Saemann:2012uq,Saemann:2017zpd,Saemann:2017rjm} (see also~\cite{Saemann:2010cp}), the M-theory lift of monopoles~\cite{Howe:1997ue}. The theory conjecturally describes the tensor branch of certain $(1,0)$-theories arising e.g.~in F-theory.
        
        \paragraph{M5-branes.} 
        A particularly interesting six-dimensional superconformal field theory is the $(2,0)$-theory~\cite{Witten:1995zh,Seiberg:1996vs}, which is supposed to be the replacement of maximally supersymmetric Yang--Mills theory in the low-energy effective description of stacks of flat D-branes for stacks of flat M5-branes, cf.~\cite{Dasgupta:1995zm,Witten:1996hc}. This theory is expected to be a local quantum field theory~\cite{Seiberg:1996qx}, and its observables are Wilson loops~\cite{Ganor:1996nf,Corrado:1999pi,Chen:2007ir}. This suggests that this theory is a quantum higher gauge theory. A classical Lagrangian has not been found, and there are a number of arguments against its existence, see~\cite{Lambert:2019khh,Saemann:2019leg} for reviews of these. Some aspects, however, seem visible, e.g.~the expected BPS states seem to agree with the non-Abelian versions of self-dual strings found in the $(1,0)$-theory above. Also, other obstacles to a Lagrangian formulation can be considered as overcome by adjusted connections on principal $2$-bundles. A description of M5-branes in terms of higher Chern--Simons theory was given in~\cite{Fiorenza:2012tb}.
        
        Under the assumption of hypothesis $H$ (cf.~\cref{sec:related_areas}), it has been shown that a single M5-brane leads to a string structure~\cite{Fiorenza:2020hiq}. At the local and infinitesimal level, this implies essentially to the field content of~\eqref{eq:1_0_theory}. Similarly, the field content for a single heterotic M5-brane has been classified in~\cite{Fiorenza:2020xpx}. For further closely related and significant work, see also~\cite{Fiorenza:2019usl,Fiorenza:2019ckz,Fiorenza:2019ain,Fiorenza:2015gla}. For a useful discussion of the emergence of higher structures in M-theory, see~\cite{Fiorenza:2019ckz}.
        
        \subsection{Higher lattice gauge theories}
        
        In lattice gauge theory~\cite{Wegner:1971app,Wilson:1974sk}, the spacetime continuum is replaced by a discrete lattice. This is achieved by assigning the group elements encoding a discrete parallel transport to the edges of an $n$-dimensional lattice. This replaces the path integral by a finite-dimensional integral and opens up the theories to numerical studies.
        
        There is an evident generalisation to higher gauge theories in which one assigns the $n$-cells of a higher parallel transport to the $n$-dimensional boundaries of a higher dimensional lattice. In particular, for a gauge $2$-group given by a crossed module of Lie groups $(\sfH\xrightarrow{~\partial~}\sfG,\acton)$, one assigns to each edge an element in $\sfG$ and each face an element in $\sfH$.
        
        Three of the earliest papers on lattice higher gauge theory are~\cite{Nepomechie:1982rb}, which proposes a lattice action for a non-Abelian 2-form, ~\cite{Omero:1982hp}, which studies the higher Stueckelberg model, cf.~\cref{ssec:Stueckelberg}, and~\cite{Frohlich:1982gf}, which refers to higher gauge theories as \uline{hypergauge theories}.
        
        A more general definition and a list of interesting examples of higher lattice gauge theory were presented in~\cite{Pfeiffer:2003je}, see also~\cite{Girelli:2003ev}, in which a higher-dimensional parallel transport is developed which is similar to the one of~\cite{Schreiber:0802.0663}. A somewhat simpler \uline{lattice gerbe model} assigning both Abelian and non-Abelian degrees of freedom to faces was studied in~\cite{Lipstein:2014vca}; see also~\cite{Bochniak:2021ece} for another model. 
        
        For a mathematical study of higher lattice gauge theories (which are, in particular, related to Yetter's homotopy $2$-type topological quantum field theory~\cite{Yetter:1993dh}), see~\cite{Bullivant:2016clk,Bullivant:2017sjz} as well as~\cite{Delcamp:2018wlb,Bullivant:2019tbp}. For work linking higher lattice gauge theories to non-linear sigma-models, see also~\cite{Zhu:2018kzd}.
        
        In all of the above models, the higher gauge theories studied are fake-flat, and we are not aware of any work discussing adjusted higher lattice gauge theories.
        
        \subsection{Other field theories}
        
        This section summarises a few other forms of higher gauge theory which are not directly related to any of the above topics.
        
        Having categorified the structure group, it is reasonable to study also the case of categorified base spaces. This has been done, albeit to a limited extent, see e.g.~\cite{Ritter:2015zur} for higher gauge theory on Courant algebroids and~\cite{Sharpe:2019yag} and references therein for sigma models with stacks (in particular, Lie groupoids) as target spaces.
        
        In the context of gravity, there is a formulation of teleparallel gravity as a higher gauge theory~\cite{Baez:2012bn} with the higher gauge group arising from the evident action of the Poincar\'e group on a copy of flat spacetime. In this description, however, parts of the gauge transformations of the principal $2$-bundle had to be excluded. Using such restricted gauge transformations, one can also reformulate M2-brane models as a higher gauge theory~\cite{Palmer:2013ena}.
        
        The kinematic data of higher gauge theory such as Wilson surfaces can be regarded as partition functions for non-linear sigma models~\cite{Zucchini:2022svl,Zucchini:2022avl}.
        
        Finally, we note that also higher versions of Abelian Dijkgraaf--Witten theories have been defined~\cite{Monnier:2015qxa}.
        
        \section{Related areas}\label{sec:related_areas}
        
        There are a number of research topics closely related to higher gauge theory. The following gives a concise summary of some of them, together with an explanation of the relation.
        
        \paragraph{Hypothesis $H$.} 
        Hypothesis $H$\footnote{$H$ may stand for homotopy.} is an attempt to define M-theory non-perturbatively, unifying the zoo of string theories in different approximations and reproducing the dualities between them, see~\cite{Alfonsi:2023pps} in this volume for a short review. Hypothesis $H$ conjectures that in M-theory, the curvature $4$-form of the $C$-field and its Hodge-dual form an element in some non-Abelian generalised cohomology theory valued on $S^n$ called $J$-twisted cohomotopy theory understood as Dirac charge quantisation~\cite{Sati:2013rxa}. This assumption implies a list of expected anomaly cancellation conditions~\cite{Sati:2020cml}. In particular, the rational approximation of $S^4$~\cite{Fiorenza:2019ckz} gives the correct equations of motions for the $C$-field in $11$-dimensional supergravity, and the M5-brane is also discussed in rational approximation in~\cite{Fiorenza:2019cqe,Fiorenza:2020xpx}. See also~\cite{Sati:2021uhj} for further work.
        
        \paragraph{Higher-form symmetries.} 
        Conventional ($0$-form) global symmetries induce powerful structural features in quantum field theory, such as  Ward identities. Their anomalies are renormalisation group invariant and, so, shine a light on non-perturbative phenomena. The seminal paper~\cite{Gaiotto:2014kfa} stimulated rapid advances in  generalisations of conventional global symmetries to \uline{higher-form global symmetries}, which similarly entail important and otherwise obscure insights. See~\cite{Sharpe:2015mja} for an early comment on these symmetries from the mathematical perspective on higher structures, and e.g.~\cite{Cordova:2022ruw,Gomes:2023ahz,Schafer-Nameki:2023jdn,Brennan:2023mmt,Luo:2023ive,Bhardwaj:2023kri} and the references therein for a glimpse of the literature.  
        
        For an $n$-dimensional theory, a conventional $0$-form symmetry operator, that acts on point-like local operators, can be generated by exponentiating the associated Noether charge, which in turn is given by the integral of the Noether $1$-form current over an $(n-1)$-dimensional manifold. This is a topological operator in the sense that does not depend on continuous deformations of the $(n-1)$-dimensional manifold. Similarly, a topological higher $p$-form symmetry operator can be generated by the integral of a higher $(p+1)$-form Noether current over an $(n-1-p)$-dimensional manifold. The $p$-form symmetry operator will act on $p$-dimensional extended operators, such as Wilson $p$-branes or defects. These $p$-branes naturally couple to background $(p+1)$-form gauge potentials, cf.~the Kalb--Ramond $2$-form of string theory, that can be used to gauge (i.e.~sum over all flat curvatures) the global higher symmetry  in favourable (e.g.~'t Hooft-anomaly-free) circumstances. This will typically result in a new higher global symmetry and theory. This perspective has generated deep insights in varied contexts~\cite{Cordova:2022ruw}, from high-energy phenomenology, see e.g.~\cite{Gaiotto:2014kfa,Cordova:2022fhg,Putrov:2023jqi}, to condensed matter, see e.g.~\cite{Barkeshli:2014cna,Ji:2019jhk,Kong:2022cpy}.
        
        To make contact with higher gauge theory, as reviewed here, note that the higher $p$-form currents in question follow from global analogues of the local higher-group gauge symmetry. In particular, whenever one has a higher gauge theory, the gauge transformations with constant gauge parameters constitute a higher-form symmetry, with associated higher $p$-form conservation laws and $(p+1)$-form Noether currents; a $p$-brane that couples to a $(p+1)$-form potential of a higher-group connection will carry non-trivial higher-form charges that obey these higher-form conservation laws. Starting from the higher gauge theory and constructing the corresponding global higher symmetry operators effectively  proceeds in the opposite direction to the usual picture of gauging a known (higher) global symmetry, generating a new global higher symmetry. Consequently, the  higher symmetry is encoded in the initial higher gauge theory, and the topological symmetry operators are given by higher holonomies of the higher gauge fields as in \cref{sec:intro}. For $p>0$, the higher topological symmetry operators will be Abelian for physical and mathematical reasons. In particular, physically, codimension $p+1>1$ defects can be continuously moved past one another. Mathematically, consistent parallel transport requires fake-flat higher connections, which renders them locally quasi-isomorphic to Abelian higher connections. It is interesting to note, however, that the fake-flatness condition can be relaxed via adjustment, cf.~\cref{ssec:local_adjustments}. This may have implications for higher global symmetries. 
        
        Note that the higher gauge groups considered above are, indeed, always higher groups as in \cref{sec:higher_groups}. However, the composition of the associated (or otherwise) topological global symmetry operators may not obey any (higher) group law. This leads to another major recent development: non-invertible symmetries. The existence of non-invertible $0$-form symmetries is well-known in two dimensions~\cite{Frohlich:2004ef,Feiguin:2006ydp,Fuchs:2007tx,Frohlich:2009gb, Bhardwaj:2017xup,Chang:2018iay,Thorngren:2019iar,Komargodski:2020mxz,Thorngren:2021yso,Cordova:2022ieu}. They are non-invertible in the sense that the composition of two topological symmetry operators may not obey a (higher) group law. Instead, they will form some fusion category, see e.g.~\cite{Etingof:2015aa}, as is well-known in the context of two-dimensional rational conformal field theory, integrable field theory, and topological quantum field theories~\cite{Zamolodchikov:1978xm,Moore:1988qv,Moore:1989yh,Fuchs:2003id,Bhardwaj:2017xup}. More recently, non-invertible symmetries have been vastly generalised to many exotic (e.g.~non-Abelian anyons in three dimensions, cf.~\cite{Kaidi:2021gbs}) and familiar (e.g.~Maxwell theory, cf.~\cite{Gaiotto:2014kfa}) cases in higher dimensions~\cite{Heidenreich:2021xpr,Choi:2021kmx,Kaidi:2021xfk,Apruzzi:2022rei, Bhardwaj:2022yxj,Choi:2022zal,Choi:2022jqy,Bashmakov:2022jtl,Lin:2022xod,Bartsch:2022mpm, GarciaEtxebarria:2022vzq,GarciaEtxebarria:2022jky,Bartsch:2022ytj}. This emerging paradigm has generated insights in a variety of contexts, reflecting the ubiquity of these higher symmetry structures: anyons, dualities and symmetry protected phases in condensed matter, e.g.~\cite{Barkeshli:2014cna,Ji:2019jhk,Kong:2022cpy}, the transition between confinement and deconfinement in relativistic gauge theories, e.g.~\cite{Gaiotto:2014kfa,Gaiotto:2017yup}, possible origins of neutrino masses~\cite{Cordova:2022fhg}, and five- and six-dimensional superconformal field theories arising in string/M-theory, e.g.~\cite{DelZotto:2015isa, DelZotto:2022fnw, Cvetic:2022imb, Heckman:2022muc, Bhardwaj:2022yxj,Bashmakov:2022jtl}, to name but a few. 
        
        \paragraph{Higher Berry connections.}
        An even more direct application of higher gauge theory arises in condensed matter via tensor, or higher, Berry connections~\cite{Palumbo:2018fvw,Palumbo:2021xav,Zhu:2021ruu}. The Bloch states $|u\rangle$ corresponding to a quantum particle on a lattice typically have a local (in momentum space) Abelian gauge symmetry with a conventional Abelian $1$-form Berry connection $A=\langle u|\rmd|u\rangle$. Generalising, one can construct from the Bloch states a $2$-form Abelian Berry connection $B$, i.e.~a gerbe, and higher $p$-form analogues~\cite{Palumbo:2018fvw}. These constructions elucidate, or re-articulate, known topological invariants, such as the first Chern number of two-dimensional Chern insulators as the two-dimensional Zak phase of a higher Berry connection and the winding number of three-dimensional topological insulators in terms of the Dixmier--Douady class of the associated Berry gerbe. The $2$-form Berry connections have been shown to have experimentally detectable consequences in synthetic condensed matter systems~\cite{Chen:2020ggw}, such as synthesised tensor monopoles~\cite{Tan:2020yci,Ding:2020xrr}, and admit numerous generalisations and applications, including non-Abelian higher connections and dualities~\cite{Palumbo:2019tmg,Zhu:2020sry,Dubinkin:2020kxo,Palumbo:2021xav,Zhu:2021ruu}.
        
        \paragraph{Axions.} 
        Higher gauge theories may also have to say something about axions. The \uline{axion} is a conjectured particle or field which was originally postulated as a solution to the \uline{strong CP problem}, a rather cosmetic problem of the standard model that consists of the fact that a particular free parameter is very close to zero. Also, the axion has in recent years become the leading particle candidate to provide the missing \uline{dark matter} in the cosmos, see~\cite{Chadha-Day:2021szb} for a detailed account.
        
        In the context of string theory, axions are readily found, see e.g.~\cite{Svrcek:2006yi}; direct phenomenological models, however, lead to tensions with experimental observations. 
        
        Concretely, terms such as
        \begin{equation}\label{eq:theta-term}
            \frac{\theta}{32\pi^2}\int\rmd^4x\,\inner{F}{F}
        \end{equation}
        can be included into the standard model where $F$ is the quantum chromodynamics (QCD) field strength and $\theta$ measures the Yang--Mills instanton number and $\inner{-}{-}$ is a suitable metric on the QCD gauge Lie algebra. A solution of the strong CP problem would have to explain why $\theta$ is small. 
        
        Terms of the form~\eqref{eq:theta-term} are naturally produced from actions containing adjusted higher curvatures. In particular, the adjustment in the field strengths often lead to topological terms in the action functional that contain products $\inner{F}{F}$, multiplied by further higher form potentials. Restricting them to their zero modes, e.g.~in the usual Kaluza--Klein reduction procedure, produces precisely terms of the form~\eqref{eq:theta-term}.
        
        As an example, consider the six-dimensional $\caN=(1,0)$-supersymmetric higher gauge theory discussed in \cref{ssec:6dCFT}. In the formulations of~\cite{Samtleben:2011fj} or~\cite{Saemann:2017zpd}, this theory contains a term of the form $B\wedge\inner{F}{F}$ in the Lagrangian, where $B$ is a $2$-form potential and $F$ the $2$-form curvature of a $1$-form potential. Considering a Kaluza--Klein reduction from six to four dimensions with $B=\frac{\theta}{32\pi^2}$ constant in the reduced dimensions produces the term~\eqref{eq:theta-term} in the reduced action.
        
    \end{body}
    
\end{document}